\documentclass[useAMS,usenatbib]{mn2e}
\usepackage{color}
\usepackage{array} 
\usepackage{amsmath}
\usepackage{graphicx}
\usepackage{epstopdf}
\usepackage[labelfont=bf,labelsep=period,justification=raggedright,singlelinecheck=false,font=small]{caption}
\usepackage{wrapfig}
\usepackage{float}
\usepackage{amssymb}
\usepackage[normalem]{ulem}
\usepackage{hyperref}
\usepackage{undertilde}
\usepackage[T1]{fontenc}
\usepackage[utf8]{inputenc}
\def\apj{ApJ}
\def\aap{A\& A}
\def\apjl{ApJL}
\def\apjs{ApJS}
\def\mnras{MNRAS}
\def\jcap{JCAP}

\def\aj{AJ}

\def\nat{Nature}
\def\prd{Phys. Rev. D}

\newcommand{\rvir}{R_{\rm vir}}

\newcommand{\msun}{{\rm M}_{\odot}}

\restylefloat{table}

\begin{document}

\title[The Impact of Baryonic Physics on the Structure of Dark Matter Halos]{The Impact of Baryonic Physics on the Structure of Dark Matter Halos: the View from the FIRE Cosmological Simulations}

\author[T. K. Chan et al.]
  {T. K. ~Chan$^1$ \thanks{Email: (TKC) tkc004@physics.ucsd.edu} , D. Kere{\v s}$^1$ \thanks{Email: (DK) dkeres@physics.ucsd.edu},
  J.~O{\~n}orbe$^{2}$, P.F.~Hopkins$^3$, A.L.~Muratov$^1$, \newauthor
   C.-A.~Faucher-Gigu\`{e}re$^4$ \& E.~Quataert$^5$ \\
  $^1$Department of Physics, Center for Astrophysics and Space Sciences, University of California at San Diego,\\ 9500 Gilman Drive, La Jolla, CA 92093, USA\\
  $^2$Max-Planck-Institut fuer Astronomie, Koenigstuhl 17, 69117 Heidelberg, Germany\\
  $^3$TAPIR, Mailcode 350-17, California Institute of Technology, Pasadena, CA 91125, USA\\
  $^4$Department of Physics and Astronomy and CIERA, Northwestern University, 2145 Sheridan Road, Evanston, IL 60208, USA\\
  $^5$Department of Astronomy and Theoretical Astrophysics Center, University of California Berkeley, Berkeley, CA 94720, USA}

\maketitle

\begin{abstract}
We study the distribution of cold dark matter (CDM) in cosmological simulations from the FIRE (Feedback In Realistic Environments) project, for $M_{\ast}\sim10^{4-11}\,M_{\odot}$ galaxies in $M_{\rm h}\sim10^{9-12}\,M_{\odot}$ halos. FIRE incorporates explicit stellar feedback in the multi-phase ISM, with energetics from stellar population models. We find that stellar feedback, without ``fine-tuned'' parameters, greatly alleviates small-scale problems in CDM. Feedback causes bursts of star formation and outflows, altering the DM distribution. As a result, the inner slope of the DM halo profile ($\alpha$) shows a strong mass dependence: profiles are shallow at $M_{\rm h}\sim10^{10}-10^{11}\,M_{\odot}$ and steepen at higher/lower masses. The resulting core sizes and slopes are consistent with observations. 
This is broadly consistent with previous work using simpler feedback schemes, but we find steeper mass dependence of $\alpha$, and relatively late growth of cores. Because the star formation efficiency $M_{\ast}/M_{\rm h}$ is strongly halo mass dependent, a rapid change in $\alpha$ occurs around $M_{\rm h}\sim 10^{10}\,M_{\odot}$ ($M_{\ast}\sim10^{6}-10^{7}\,M_{\odot}$), as sufficient feedback energy becomes available to perturb the DM. Large cores are not established during the period of rapid growth of halos because of ongoing DM mass accumulation. Instead, cores require several bursts of star formation after the rapid buildup has completed. Stellar feedback dramatically reduces circular velocities in the inner kpc of massive dwarfs; this could be sufficient to explain the ``Too Big To Fail'' problem without invoking non-standard DM. Finally, feedback and baryonic contraction in Milky Way-mass halos produce DM profiles slightly shallower than the Navarro-Frenk-White profile, consistent with the normalization of the observed Tully-Fisher relation.
\end{abstract}


\label{firstpage}

\section{Introduction}
Cold Dark Matter with a cosmological constant ($\Lambda$CDM) is a successful cosmological model that can simultaneously explain large scale fluctuations in the cosmic microwave background and the large-scale structure of the universe that forms out of these fluctuations at much later time \citep{Sper07, Spri05b}. However, on much smaller scales, within dark matter halos that host observed galaxies, there are indications that the distribution of dark matter is inconsistent with the simplest prediction of the cold dark matter paradigm. The most obvious and most studied disagreement is in density profiles of dark matter halos inferred from observations of dwarf and low surface-brightness galaxies. While observed slopes are relatively flat (central density slope $\alpha\sim 0$, where central density $\propto r^\alpha$) \citep{Salu00, Swat03, Gent04, Spek05, THINGS, deBl08, Oh11} simulated cold dark matter halos are cuspy ($\alpha\sim -1$) \citep{Flor94,Moor94,NFW}. This problem is known as the {\it cusp/core } problem.

To address this problem, various modifications of dark matter properties have been proposed to erase the steep central regions and produce a core-like density profile. Examples include warm dark matter (WDM), whose free streaming can suppress the small scale structure \citep{Abaz06,Duns11,Love12} and self-interacting dark matter (SIDM) whose interaction can substantially affect the central density profile of halos \cite[e.g.][]{Yosh00, Burk00, Koch00, Dave01,Elbe15}. There is still no consensus on whether these modifications can solve the problem and satisfy all observational constraints: these modifications can produce serious problems on their own. For example \cite{Macc12a} found that in order to produce dark matter cores as large as those seen in observed dwarf galaxies, the warm dark matter would also prevent the formation of dwarf galaxies. Simple SIDM models \citep{Carl92,Mach93,deLa95} were shown to violate observations of central regions of galaxy clusters \citep{Mira02, Yosh00} that are found to be denser and more elliptical than SIDM would predict.  However, recent SIDM models that take into account more accurate observational constraints and the effects of baryons offer promising explanation of the problem without violating any known observational constraints \citep{Roch13,Pete13,Kapl14,Elbe15}.

Initial problems with the SIDM have motivated more complex models such as velocity-dependent SIDM \citep{Yosh00,Loeb11,Macc12a}. However, SIDM with a simple power-law velocity dependence will not be able to create a core and, at the same time, produce stable halos of dwarf galaxies over a Hubble time \citep{Gned01}. \cite{Loeb11} proposed SIDM with a Yukawa potential, which has a nontrivial velocity-dependence that is effective at producing cores in dwarf galaxies without adverse effects on clusters of galaxies. Cosmological simulation of a Milky Way-mass halo with this SIDM showed that realistic cores can be formed in sub-halos expected to host dwarf galaxies \citep{Voge12}. However, this model requires more free parameters for the velocity dependence and it is not yet known whether it can reproduce the correct halo abundance and mass distribution.

Before concluding that simple cold dark matter models must be modified,, we must also examine the effects of baryons on the distribution of dark matter within halos. Baryons are not only what is actually observed in galaxies, but baryonic effects at the halo center can, in principle, also play a role in shaping the dark matter profiles \citep{Blum86,Nava96,ElZa01,Gned04,Read05,Gove10,Pena12,Gove12,Pont12,Macc12b,Teys13,DiCi14,Pont14}.

\cite{Nava96} and \cite{Read05} used N-body simulations to model a sudden removal of a large baryonic component via supernova-driven winds (represented as a change in the external potential) in a dwarf galaxy with an initially cuspy dark matter halo. They showed that such mass removal leads to formation of a dark matter core. An alternative mechanism was proposed by \cite{Mash06},  who showed that bulk motion of gas within forming galaxies leads to significant gravitational potential changes which can also redistribute dark matter and reduce its central density. Dynamical effects between baryons and dark matter during the halo formation were also suggested as a mechanism that could modify dark matter density profiles \citep[e.g.][]{ElZa01,Toni06,Roma08,DelP09}.

Recently \cite{Gove10,Gove12} used cosmological ``zoom-in simulations'' with baryons, cooling, star formation and supernovae feedback to show that outflow episodes in dwarf galaxies can turn the central dark matter cusps into cores.
Strong supernova-driven outflows from clustered star formation in the inhomogeneous interstellar medium (ISM) resulted in a decrease of the dark matter density within central kiloparsec to less than half of what it would otherwise be in a halo of this mass ($m_{\rm dm} \approx 10^{10} \msun$). 
\cite{Pont12} further clarified this density flattening mechanism: a quick change in gravitational potential due to gas outflow can effectively inject energy into dark matter orbits and (typically after many outflow episodes) flatten the central dark matter profile. They showed that the repeated changes of gravitational potential on timescales shorter than $t_{\rm dyn}$ during $2 < z < 4$ can significantly flatten cuspy dark matter profiles. 

\cite{Broo12} showed that a large fraction of the gas that is expelled returns via a large-scale galactic fountain \citep[see also][]{Oppe10} to form stars at later times: this greatly increases the chance of outflows from the inner regions and further helps in core formation. Other, non-cosmological simulations with strong SNe feedback also showed that it is possible to form cores in dwarf galaxies owing to bursty star formation that removes large quantities of gas during bursts \citep{Teys13}. 

On the other hand, idealized simulations by \cite{Gned02},\cite{Ogiy11}, and \cite{Garr13} focused on time evolution, supernovae energy requirements and mass ejection frequency in idealized models and argued that SNe driven feedback is not efficient enough to form cores at the observed level. However, recent cosmological simulations have had more success.  \cite{Mada14} suggest that earlier mass removal can lower the energy requirements for core formation and \cite{Onor15} showed that late star formation, after the early epoch of cusp building, is particularly efficient at utilizing stellar feedback to remove dark matter.

There are two other related ``problems'' with structural properties of CDM halos. One is the lack of very steep central profiles in relatively massive disk galaxies. Cuspy NFW profiles  \citep{NFW} are expected to be even steeper within baryon dominated galaxies owing to the contraction of dark matter caused by the central concentration of baryons \citep{Blum86}.
The distribution of matter in galaxies effectively determines the Tully-Fisher relation \citep[e.g.][]{Tull76,Dutt07}. However, given the observed distribution of baryons, contraction of an NFW halo would result in circular velocities too high at a given luminosity/mass. This motivated several authors to suggest that dark matter does not undergo contraction or is perhaps even expanded from the original cuspy NFW profile \citep{Dutt07}. While this could be interpreted as suggesting a problem with the currently favored CDM model, stellar feedback is also able to effectively ``expand'' the dark matter distribution even in Milky-Way mass halos \citep{Macc12b}.

The second problem is the so called ``Too Big To Fail" problem \citep{Boyl11}. In the Milky Way, satellites have significantly lower dark matter densities in the inner few hundred parsecs than the corresponding sub-halos in CDM only simulations without baryons. Alternatively, massive sub-halos whose inner densities are high, never formed galaxies. Similar problems also exist in other dwarf galaxies in the Local Group \citep{Garr14,Papa15}.
There are hints that feedback can help solve this issue along with the cusp/core problem \citep{Mada14,Broo14, Onor15}, although proper statistics are still lacking.

It is becoming clear that the bursty nature of stellar feedback in galaxies can modify the inner regions of dark matter halos. However, in general, most simulations used to study this problem so far used crude and often unphysical implementations of stellar feedback. One might worry that this could impact the effect of stellar feedback. Most of the cosmological simulations used to address the cusp/core issue simply turn off cooling from the gas heated by supernovae ejecta until such gas escapes galaxies  \citep{Gove10, Gove12, Pont12, Macc12b, DiCi14}. The delayed cooling is unphysically long and results from a misinterpretation of the standard supernova remnant results \citep{Mart15}. In addition, most simulations include only supernovae feedback while other stellar feedback mechanisms are ignored or implemented crudely: e.g. radiation pressure, cosmic rays, and photo-heating are often approximated with pure thermal energy input and additional freely-adjustable parameters \citep{Macc12b,DiCi14}. 

Furthermore, the particle mass resolution used in some previous studies was insufficient to properly resolve the observed core sizes. Low resolution may hinder the investigation of central density profiles on small scales owing to two-body relaxation effects \citep{Powe03}. In Appendix \ref{relimit},  we show the relation between the convergence radius and particle mass that should be used to estimate resolved scales in different simulations.

To study the {\it cusp/core} problem in a complex cosmological environment we use simulations from the Feedback in Realistic Environments (FIRE) project \citep{FIRE}\footnote{Project website: http://fire.northwestern.edu/} (H14), supplemented by four new dwarf galaxy simulations. Our simulations use physically motivated stellar feedback, in which energy and momentum input are based on stellar population synthesis models alone, with no adjusted parameters. In additions to supernovae energy and momentum we include radiation pressure, stellar winds, photo-ionization and photo-electric heating processes. In H14 we show that the $M_{*}$-$M_{\rm h}$ relation in FIRE is in reasonable agreement with observations, for galaxies residing in halo masses $M \lesssim 10^{12}\msun$. This result is sensitive to the feedback physics: simulations with supernovae alone fail to reproduce the correct relation, unless additional feedback processes are also incorporated. Overall, stellar feedback in FIRE simulations results in bursty star formation histories followed by strong outflow episodes \citep{Mura15} that can affect matter distribution within galaxies.

Our simulations are among the highest resolution cosmological ``zoom-in" simulations to date evolved down to $z=0$ with full baryonic physics. In addition to the advantages in implemented physics and resolution, we adopt the P-SPH ``pressure-entropy'' formulation of SPH \citep{Hopk13}, which includes a large number of numerical improvements relative to previous SPH studies, which together significantly improve the treatment of cooling and multi-phase fluid mixing, and reduce the well-known discrepancies between SPH and grid-based codes.

In this paper we study halos with masses $10^9< M_{\rm h}/\msun< 10^{12}$ with full feedback and their dark matter only analogs, which enables us to directly compare their dark matter distributions. We find results in broad agreement with previous work, but with some important differences. We find that stellar feedback affects all of the systems we study but large cores develop only in the halo mass range of $\sim 10^{10}\msun$ to a few $\times 10^{11}\msun$.
Furthermore we show that cores change over time, and that progenitors of massive galaxies once had more prominent cores. We demonstrate how bursty star formation and related feedback correlate with changes in dark matter halos and show that feedback effectively cancels the effects of adiabatic contraction. Finally we discuss consequences of our results for the cusp/core issue, the Tully-Fisher relation, the ``Too Big To Fail" problem and indirect dark matter detection. 
 
We find encouraging trends that have the potential to solve most of the apparent small scale problems of the CDM paradigm. 

The paper is organized as follows: \S\ref{numerical} includes a brief description of the code and implemented stellar feedback as well as the set-up of the simulations. In \S \ref{results}, we show the dark matter density profiles and their time evolution. In \S \ref{haloexpansion_tully} we study the effects of stellar feedback on the expected contraction of dark matter and on the Tully-Fisher relation. In \S \ref{discussion}, we compare our results with previous work, discuss the implications and propose directions for further investigation.

\section{Simulations}
\label{numerical}

\subsection{Simulation Code}
The simulations in this work were run with the newly developed GIZMO code \citep{Hopk14} in a fully conservative, pressure-entropy based smoothed particle hydrodynamic (P-SPH) mode \citep{Hopk13}. 
P-SPH eliminates artificial surface tension at contact discontinuities that affects traditional density based SPH \citep{Ager07, Sait13, Sija12}. We use the artificial viscosity algorithm with a switch from \cite{Cull10} which reduces viscosity to close to zero away from shocks and enables accurate shock capturing. The same higher-order dissipation switch is used to trigger entropy mixing at the kernel scale following \cite{Pric08}. Time-stepping is controlled by the limiter from \cite{Duri12}, which limits the difference in time-steps between neighboring particles, further reducing numerical errors. The gravity solver of the GIZMO code is an updated version of the PM+Tree algorithm from Gadget-3 \citep{Spri05} and uses fully conservative adaptive gravitational softening for gas \citep{Pric07}. GIZMO's softening kernel represents the exact solution of the particle mass distributed over  the SPH smoothing kernel \citep{Barn12}. 

The code performs well on standard strong shock, Kelvin-Helmholtz and Rayleigh-Taylor instabilities, and subsonic turbulence tests \citep[for more details see][]{Hopk13}. In cosmological ``zoom-in" simulations of a Milky Way size halo without outflows, the code eliminates most of the artificial fragmentation of halo gas seen in traditional SPH simulations \citep{Kauf06, Somm06, Kere09c} and increases cooling from the hot halo gas at late times \citep[e.g.][]{Kere05}, when compared to the classical SPH (Kere\v{s} et al., in preparation). Overall, the resulting halo gas properties are similar to the results from adaptive-mesh and moving-mesh simulations \citep{Ager09, Kere12, Voge12a, Nels13}.

\subsection{Baryonic Physics}

Our simulations incorporate cooling, star formation and physical stellar feedback processes that are observed to be relevant in the inter-stellar medium. Here we briefly review these components, for detailed description please see H14.

Gas follows an ionized+atomic+molecular cooling curve from $10-10^{10}$ K, including metallicity-dependent fine-structure and molecular cooling at low temperatures, and at high-temperatures ($>10^4$ K) metal-line cooling followed species-by-species for 11 separately tracked species. At all times, we tabulate the appropriate ionization states and cooling rates from a compilation of CLOUDY runs, including the effect of the photo-ionizing background. We use global ultraviolet background model from \cite{Fauc09} that heats and ionizes the gas in an ionization equilibrium approximation. We apply on-the-fly ionization corrections in denser gas to account for the self-shielding based on the local Jeans-length approximation (to determine the surface density), which provides an excellent match to a full ionization radiative transfer calculation \citep{Fauc10, Rahm13, Fauc14}.

Star formation is allowed only in dense, molecular, self-gravitating regions above $n > 10-100$ $\mathrm{cm}^{-3}$. This threshold is much higher than that adopted in most ``zoom-in" simulations of galaxy formation (the high value allows us to capture highly clustered star formation). We follow \cite{Krum11} to calculate the molecular fraction $f_{H2}$ in dense gas as a function of local column density and metallicity. We allow SF only from the locally self-gravitating molecular gas using the efficiency of 100\% per free fall time (the actual SF efficiency is feedback regulated).

Our stellar feedback model includes a comprehensive set of physical mechanisms: radiation pressure, supernovae (with appropriate momentum and thermal energy input), stellar winds, photo-ionization and photo-electric heating as described in H14. We do not tune any feedback model parameters but instead directly use the energy, momentum, mass and metal return based on the output of the STARBURST99 stellar population synthesis model \citep{Leit99}. Our feedback model is implemented within the densest interstellar-medium material, yet we do not resort to turning off cooling of supernova heated gas at any time.

\subsection{Initial conditions and zoom-in method}
We adopted a `standard' flat $\Lambda$CDM cosmology with $\Omega_0\approx 0.27$, $\Lambda\approx0.73$, $\Omega_b\approx 0.045$, $h\approx0.7$ for all runs. In order to reach the high-resolution necessary to resolve a multi-phase ISM and to properly incorporate our feedback model we use the ``zoom-in'' technique. This places maximum baryonic and dark matter resolution around the halo of interest in a lower resolution, collisionless box \citep{Port85, Katz93}.

\begin{table*}
\centering
    \begin{tabular}{lllllllll}
    \hline\hline
    Name & $M^0_{\rm h}$&$M^0_{*}$ &$R_{\rm vir}$&$R_{1/2}$& $m_{\rm b}$ & $\epsilon_{\rm b}$ & $m_{\rm dm}$ & $\epsilon_{\rm dm}$                     \\
         & [$M_{\odot}$] & [$M_{\odot}$]& [kpc] & [kpc]&[$M_{\odot}$] & [pc] & [$M_{\odot}$] &  [pc]  \\
    \hline
    	\hline
    {\bf m09}  & 2.6e9             &4.6e4           &36    &0.49                     & 2.6e2                                         & 2.0                                            & 1.3e3                                             & 29                                                                 \\
    {\bf m10}  & 7.9e9                     &  2.3e6    &50       &      0.51             & 2.6e2                                         & 2.9 & 1.3e3                                             & 29                                               \\
    {\bf m11}  & 1.4e11                   & 2.4e9      &1.4e2     &  6.9                 & 7.1e3                                         & 7.1                                            & 3.5e4                                             & 71                                                     \\
    {\bf m12v}  & 6.3e11                  &2.9e10       &2.2e2    &      1.8                  & 3.9e4                                         & 21 & 2.0e5                                             & 2.9e2                                                      \\

    {\bf m12i} & 1.1e12                       &6.1e10       &2.7e2      &4.3             & 5.7e4                                         & 20                                             & 2.8e5                                             & 1.4e2                                                                    \\
        {\bf m12q} & 1.2e12                     &2.1e10        &2.8e2    & 3.6                & 7.1e3                                         & 20 & 2.8e5                                             & 2.1e2                                                   \\
        {\bf dm09}  & 3.3e9     & -      &     39         &-                          &-  &-  & 1.6e3 & 29                                                                 \\
    {\bf dm10}  & 9.3e9           &-        &54              &-                  &-  &-  & 1.6e3 & 29  \\
    {\bf dm11}  & 1.6e11            &-        &1.4e2           &-                     &-  & - & 4.3e4                                             & 71                                                                          \\
    {\bf dm12v} & 7.7e11           &-    &         2.4e2       &-                   &-  &-  & 2.4e5                                             & 2.9e2                                                        \\
    
    {\bf dm12i} & 1.1e12            &-     &2.9e2                   &-               &-  &- & 3.4e5                                             & 1.4e2                                                                       \\
    {\bf dm12q} & 1.4e12            &-   & 3.0e2                  &-                  &-  &- & 3.4e5                                             & 2.1e2                                                                        \\
    \hline
    \hline
    {\bf m10h1297} & 1.3e10           & 1.7e7     & 62 & 1.8 &1.5e3  &4.3  & 7.3e3 & 43 \\ 
    {\bf m10h1146} & 1.6e10           &7.9e7     &         65       &2.5                  &1.5e3  &4.3  & 7.3e3 & 43 \\
    {\bf m10h573} & 4.0e10            &3.2e8    & 88                  &3.4                  &1.5e3  &10 & 7.3e3 & 1.0e2 \\
    {\bf m11h383} & 1.6e11            &4.0e9      & 1.4e2                   &7.2               &1.2e4  &10 & 5.9e4                                       & 1.0e2   \\
    {\bf dm10h1297} & 1.6e10           & -     & 66 & - &-  &-  & 8.8e3 & 43 \\ 
        {\bf dm10h1146} & 1.8e10          &-     &      68       &-                 &-  &-  & 8.8e3 & 43 \\
    {\bf dm10h573} & 4.2e10            &-    & 91                  & -                 &-  &- & 8.8e3 & 1e2 \\
    {\bf dm11h383} & 1.7e11            &-     &1.4e2                  &  -             &-  &- & 7.1e4                                       & 1.0e2                                                                     \\
    \hline
    
    \end{tabular}
    \caption{Simulation details. $M^0_{\rm h}$ and $M^0_{*}$ are the total mass and stellar mass of the largest halo in the simulation at $z=0$. $R_{1/2}$ is the radius of the region where half of the stellar mass is enclosed. $m_{\rm b}$ is the mass of a gas particle in the simulation; $m_{\rm dm}$ is the mass of a dark matter particle in the simulation. $\epsilon_b$ is the minimum gravitational smoothing length of gas; $\epsilon_{\rm dm}$ is the Plummer equivalent gravitational smoothing length of dark matter. The simulation name convention is as follows, ``mXX'' refers to the halo mass $\sim 10^{XX}\msun$. {\bf d} are the corresponding dark matter only simulations, e.g. {\bf dm09} corresponds to {\bf m09}. DM only and hydrodynamical simulations have the same initial conditions, except that the gas particles are absorbed into the dark matter particles in DM only simulations. {\bf m09}, {\bf m10}, {\bf m11}, {\bf m12v}, {\bf m12i} and {\bf m12q} are from H14, whereas {\bf m10h1297}, {\bf m10h1146}, {\bf m10h573} and {\bf m11h383} are new simulations first presented in this work.}
    \label{SIC}
\end{table*}

We consider halos with mass from $10^9$ to $10^{12}\msun$ at $z=0$ from the FIRE project \citep{FIRE}. Initial conditions of those halos are listed in Tab. \ref{SIC}. The simulations {\bf m09} and {\bf m10} are constructed using the methods from \cite{Onor14}; they are isolated dwarfs. Simulations {\bf m11}, {\bf m12q} and {\bf m12i} are chosen to match a subset of initial conditions from the AGORA project \citep{Kim13} while {\bf m12v} uses initial conditions from \cite{Fauc11a} \citep[higher resolution versions of the run first presented in][]{Kere09c}. 
In addition, we re-simulated all of these initial conditions using dark matter only N-body simulations with the same $\Omega_m$ to have a matched set of simulations with and without baryonic physics for a direct comparison. 

To improve halo mass coverage in the regime where cores are prominent, $M_{\rm h}\sim 10^{10}-10^{11}\msun$, we have also simulated additional four halos with the same FIRE code, also listed in Table \ref{SIC}. These additional simulations are first time presented in this work. Initial conditions were generated using the {\scshape MUSIC} code \citep{Hahn11}. We randomly selected halos with small Lagrangian regions for resimulation from a 40 $h^{-1}$ Mpc box. All particles within $3 R_{\rm vir}$ at $z=0$ are enclosed by the Lagrangian region to reduce contamination \citep{Onor14}. Halos {\bf m10h1297}, {\bf m10h1146} and {\bf m10h573} have masses between {\bf m10} and {\bf m11}, while {\bf m11h383} is slightly more massive. They are all isolated field dwarfs.

\subsection{Convergence Radius}
\label{resolimit}
We adopted the method described in \cite{Powe03} to calculate a conservative limit for the convergence radius of the dark matter profiles in the N-body only simulations. They found that effective resolution is related to radius where the two-body relaxation time, $t_{\rm relax}$, becomes shorter than the age of the universe $t_0$. They verified this with N-body simulations and found out that for this particular problem, well resolved regions of halos require $t_{\rm relax}>0.6t_0$. At smaller radii, even if the dynamics are locally well resolved, small N-body effects can, over a long integration time, artificially turn cusps into cores. Given the enclosed number of particles $N$ and the average density of the enclosed region $\bar{\rho}$ one can show that:
\begin{equation}
\frac{t_{\rm relax}(r)}{t_{0}}=\frac{\sqrt{200}}{8}\frac{N}{\ln N}\left(\frac{\bar{\rho}}{\rho_{\rm crit}}\right)^{-1/2},
\label{relimeq}
\end{equation}
where $\rho_{\rm crit}$ is the critical density. 
We define $r_{\rm Pow}$ as the smallest radius that fulfills $t_{\rm relax}>0.6t_0$ for the dark matter only simulations and use $r_{\rm Pow}$ to conservatively estimate where DM profiles are converged. We show the minimum particle mass required for converged profiles at $0.3-3\% R_{\rm vir}$ in Appendix \ref{relimit} and discuss implications and limitations of this convergence criterion in \S~\ref{limitations}.

\subsection{Halo Finding}
\label{AHFhalo}
We identify halos and estimate their masses and radii using the Amiga Halo Finder ({\rmfamily AHF}) \citep{Knol09}\footnote{http://popia.ft.uam.es/AHF/Download.html}. {\rmfamily AHF} uses an adaptive mesh refinement hierarchy to locate the prospective halo center \citep{Kneb01}. We use the \cite{Brya98} formulae to determine the virial over-density and virial radius, $R_{\rm vir}$, and quote the halo mass as the mass enclosed within the virial radius. We follow the main progenitor of a halo using the merger tree code included in AHF and check the growth history of each individual halo, making sure that we follow the same main progenitor. We use the histories of the main progenitors to study the time evolution of the density profile. Occasionally, AHF misidentifies the main progenitor of the halo, so sometimes $R_{\rm vir}$ temporarily decreases. We avoid this problem by searching for another halo with larger $R_{\rm vir}$ and its center within 50 kpc of the center of the main progenitor in the previous snapshot.

During ongoing mergers and for galaxies with large clumps of stars and gas, AHF adaptive centering on the highest overall density might quickly change over time and might not center on the stars. This is especially important in halos with shallow dark matter profiles and relatively shallow distributions of stars. 
To avoid this issue and have a consistent centering on the stellar component, we use two-step procedure to identify the center. First we use AHF to define $R_{\rm vir}$ and an approximate center. Within 0.1$R_{\rm vir}$ around this center we place the stellar mass volume density on a grid. We search for progressively higher overdensities which we enclose in an iso-density ellipsoid. Once this ellipsoid includes less than one quarter of the total galaxy stellar mass we stop the procedure. The center of the ellipsoid is our new halo center.
This ensures consistent centering of all profiles on the galactic stellar distribution. We tested this procedure and found that our newly defined center is closer to the DM density peak than the original AHF for cases with shallow central cores and shows less stochastic variations of the central density slope over time.

\section{Dark matter profiles and their time evolution}
\label{results}
\begin{figure*}
\includegraphics[scale=0.4]{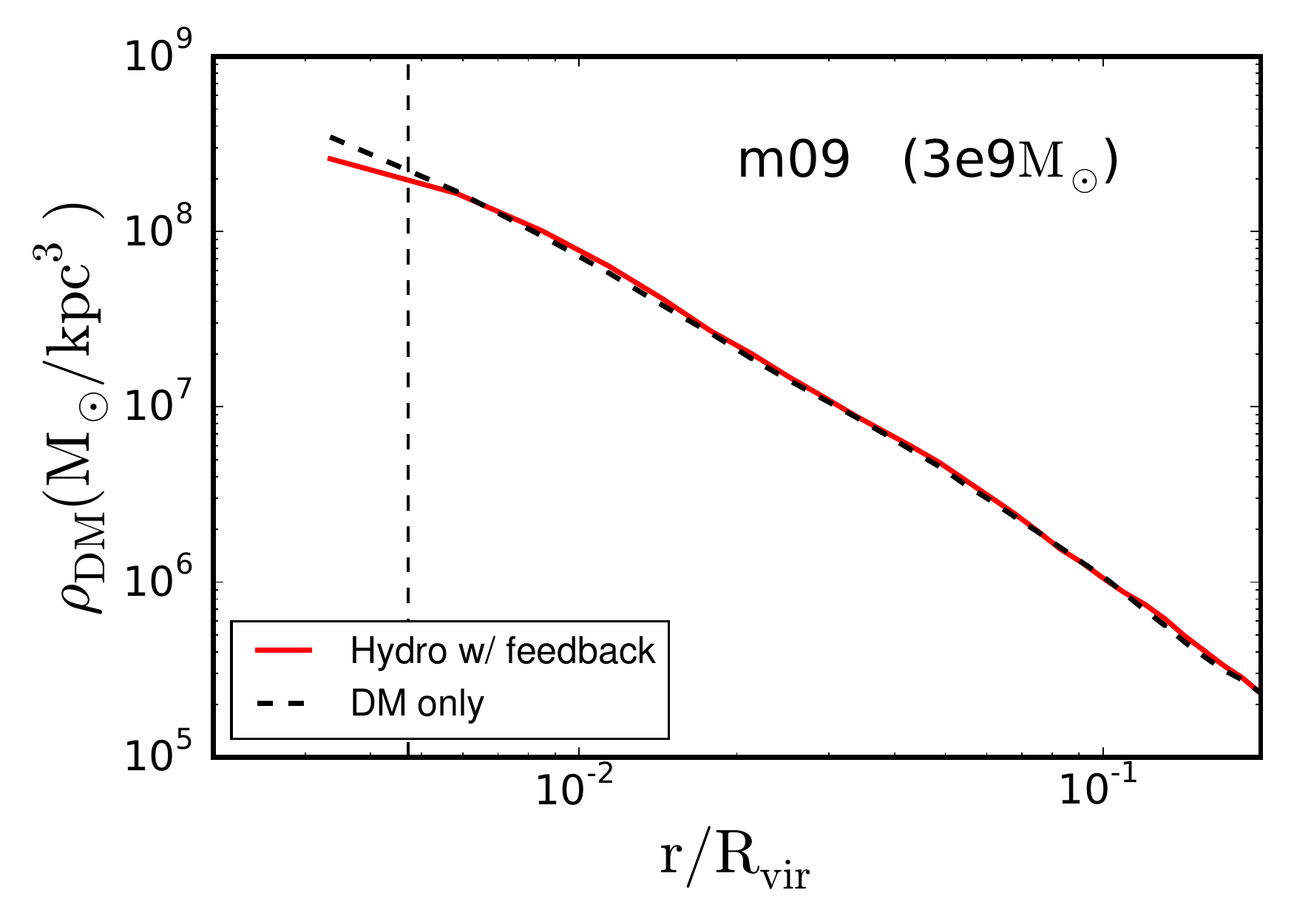}
\includegraphics[scale=0.4]{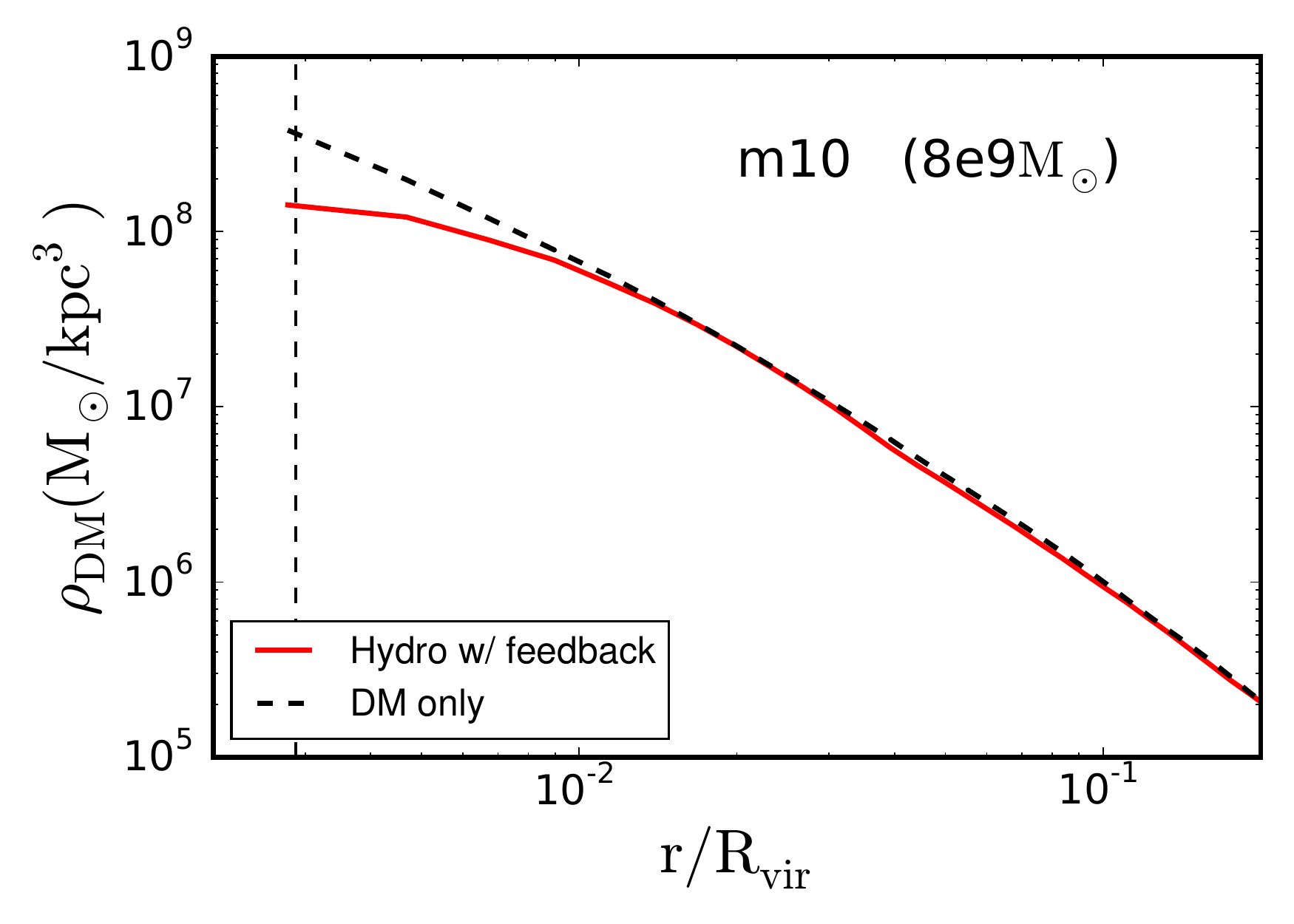}
\includegraphics[scale=0.4]{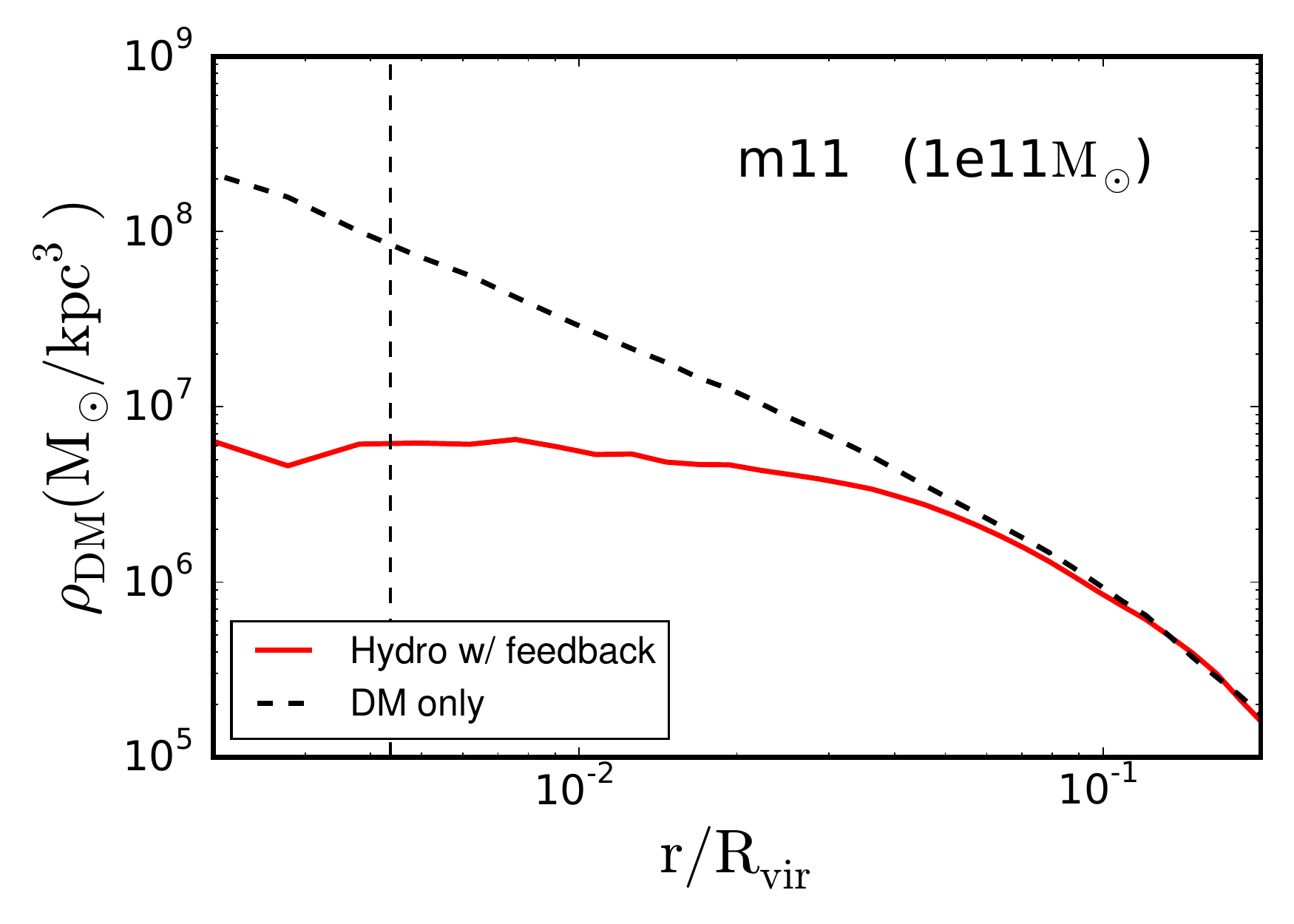}
\includegraphics[scale=0.4]{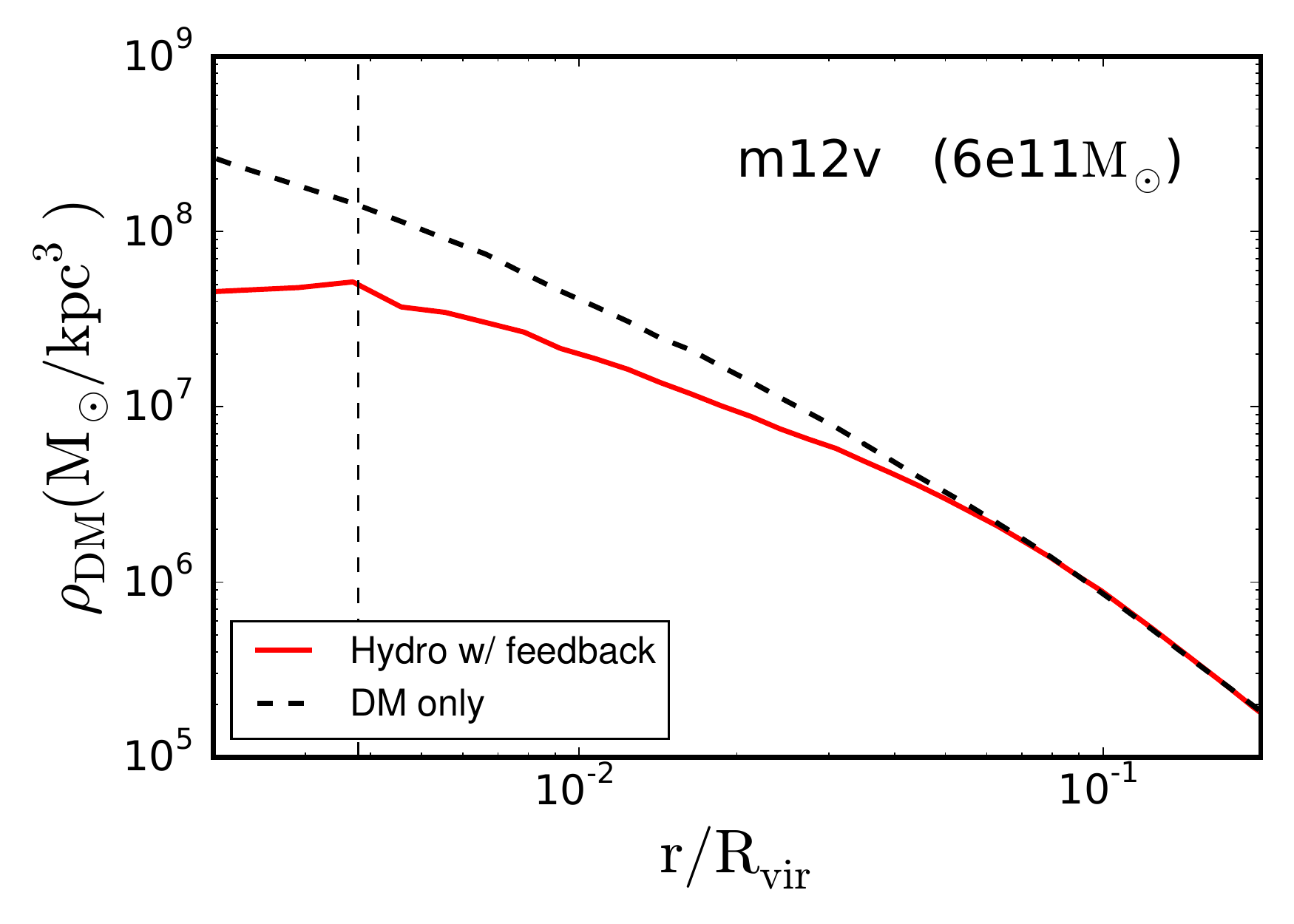}
\includegraphics[scale=0.4]{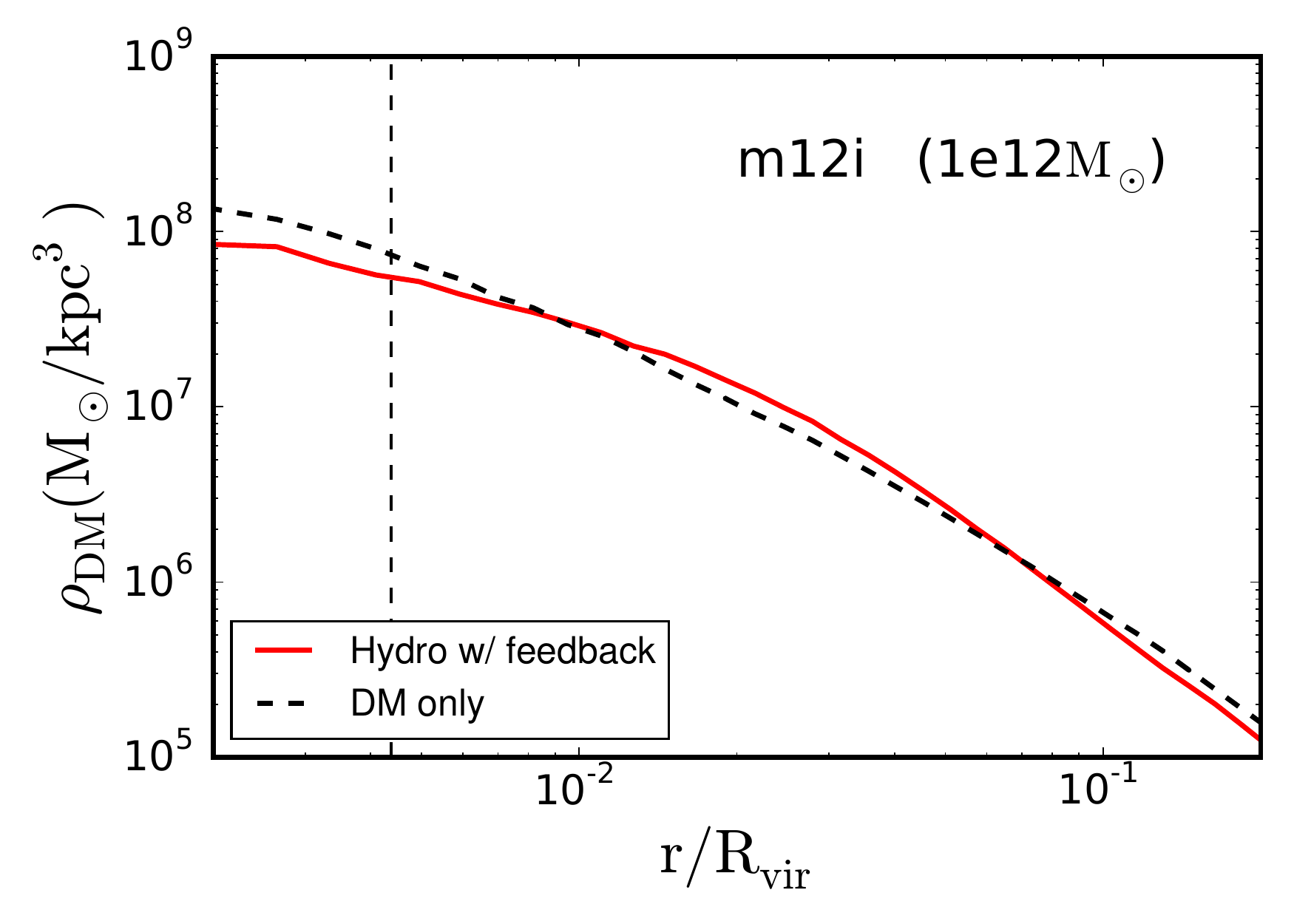}
\includegraphics[scale=0.4]{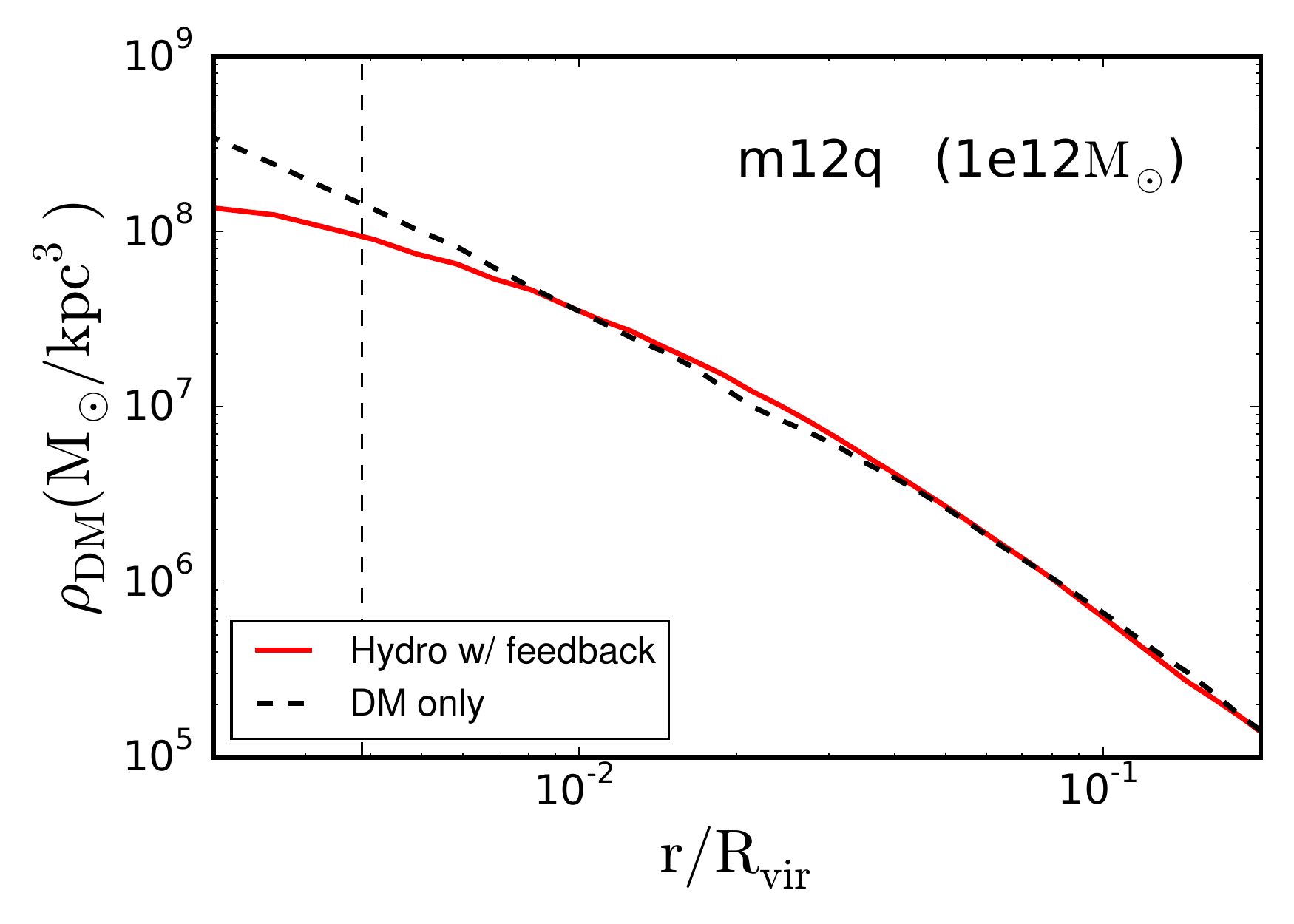}
\caption{Dark matter density profiles of halos at $z=0$. Black dashed lines represent collisionless dark matter only simulations; red solid lines represent simulations with baryons and stellar feedback. The Power radius $r_{\rm Pow}$, within which N-body relaxation effects can become important, is shown with vertical black dashed lines. The halo masses are shown in the brackets. Baryonic feedback reduces the central DM density, especially at around $M_{\rm h}\sim 10^{11}\msun$.}
\label{densitynobc}
\end{figure*}

The focus of this paper is the effect of stellar feedback on the dark matter distribution in our simulated halos. There are three major areas we explore: the relation between the halo mass and the central dark matter profile, the time evolution of the inner dark matter density, and the changes to galaxy structure caused by the effects of stellar feedback.

\subsection{Dark matter profiles}
\label{darkmatterpro}

\begin{figure}
\includegraphics[scale=0.43]{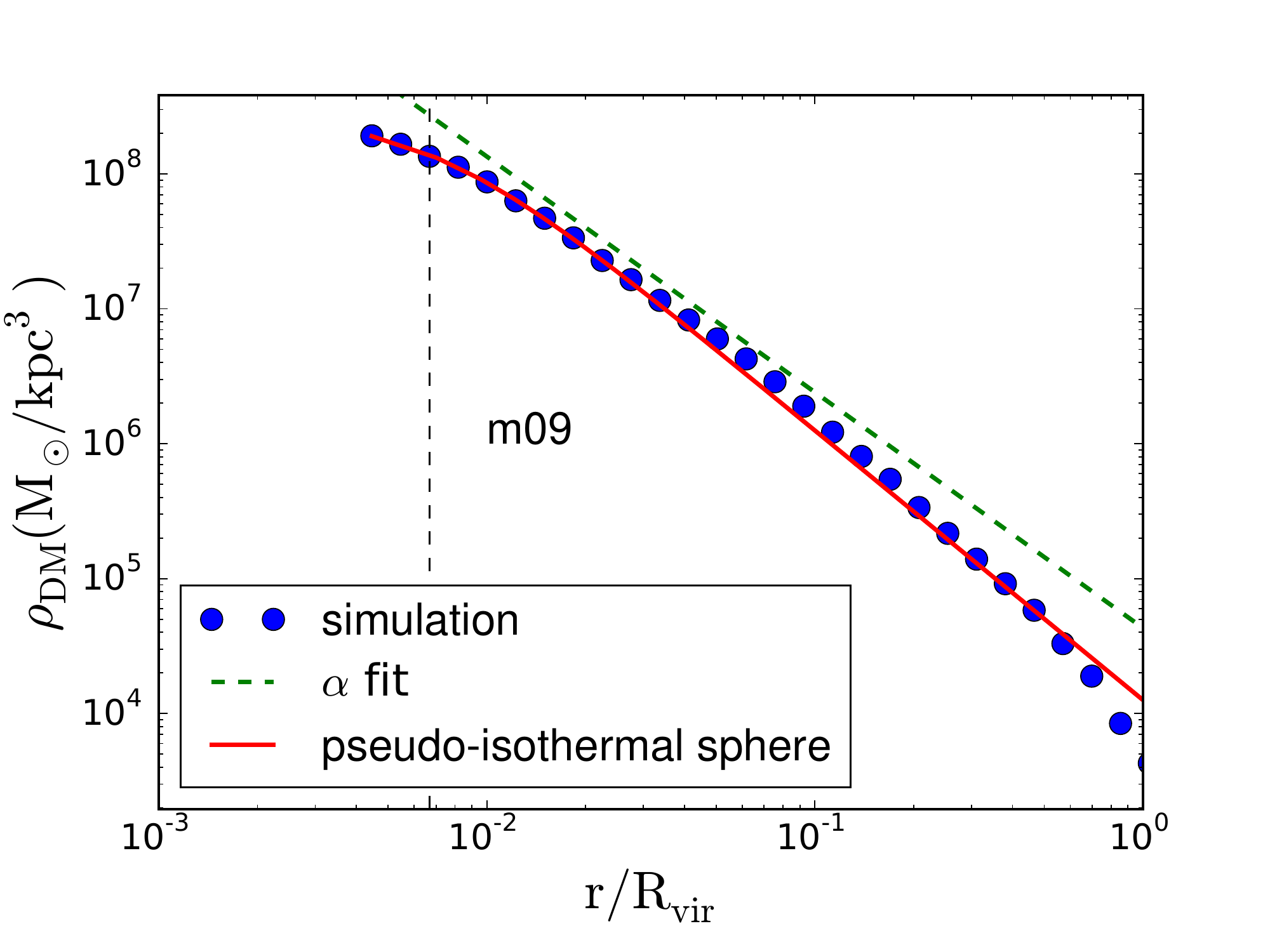}
\includegraphics[scale=0.43]{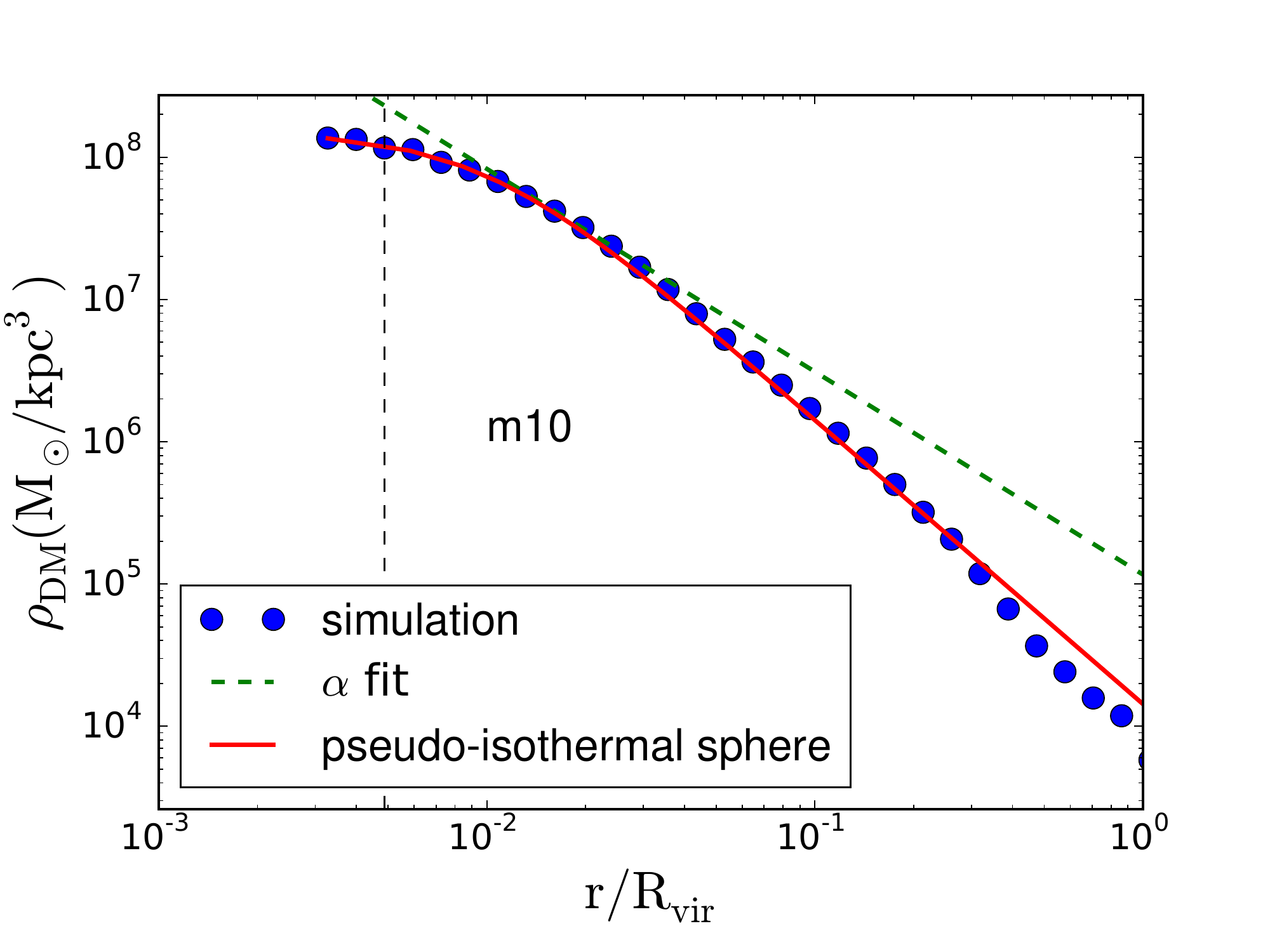}
\includegraphics[scale=0.43]{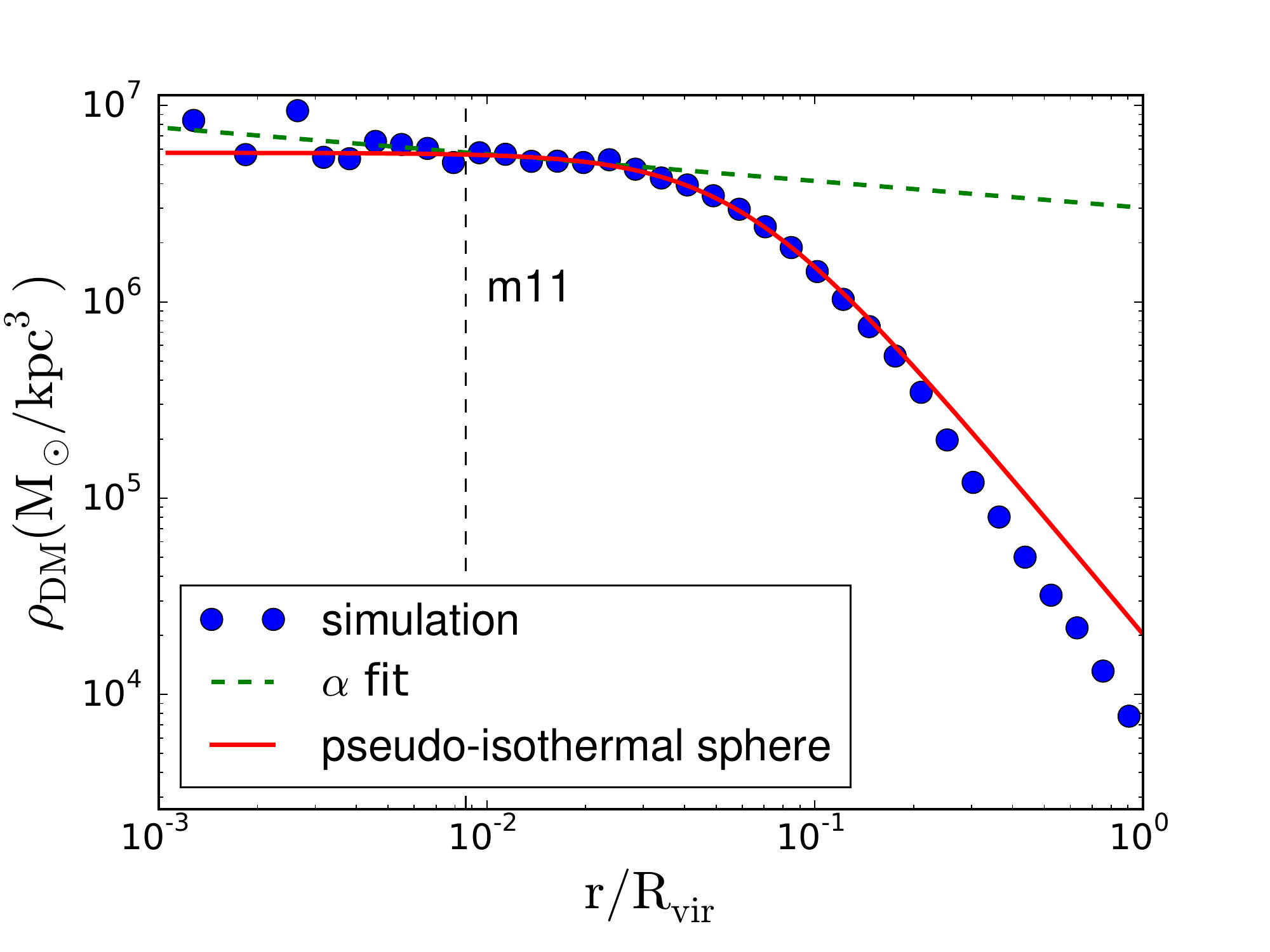}
\caption{Dark matter density profiles of halos at $z=0$ from hydrodynamic simulations (blue solid circles), fits with the pseudo-isothermal sphere (black solid line) and fit with a power-law model ($\propto r^\alpha$) at 1-2 $\% R_{\rm vir}$ (green dashed line). The black vertical line shows the convergent radius according to the Power criterion.  The pseudo-isothermal sphere is a good fit to the central regions of the simulated halos and provides a good estimate of the core sizes. }
\label{core_radius_iso}
\end{figure}

Figure \ref{densitynobc} shows spherically averaged dark matter density profiles of six simulated halos at $z=0$. We focus on the inner regions of halos, 0.002-0.2$R_{\rm vir}$  where galaxies reside and where effects of feedback are expected to be measurable. We show both the profiles from DM only simulations as well as DM profiles from simulations with baryonic physics. DM only profiles are re-normalized to account for the lower global $\Omega_{\rm DM}$ in simulations with baryons. Effects of the baryonic physics are visible in most halos to a different degree and are the largest in {\bf m11} and {\bf m12v} simulations. In {\bf m09} the DM density profile is almost the same in simulations with and without baryons, while in {\bf m10}, a small, resolved core forms in the central region.  The density profile in {\bf m11} has the largest core and the lowest central DM density. In {\bf m12v}, central density starts increasing again and the relative core size decreases, but differences in profiles are present all the way to several percent of $R_{\rm vir}$. In the two most massive halos we analyze, {\bf m12i} and {\bf m12q}, differences in central region are even smaller, although profiles are still shallower than what is expected based on N-body simulations. However in \S~\ref{haloexpansion}, we show that the effect of feedback in these halos is significant and largely cancels out the gravitational influence of baryons that is expected to steepen the profile seen in dark matter only case.  
We note that all of the plotted range is resolved with many gravitational softenings of the dark matter particles. We also show the more conservative Power convergence criterion which is typically a fraction of a percent of $R_{\rm vir}$ for all of our halos.

We quantify the effect of feedback on dark matter distribution using two parameters that are frequently estimated from observations: the central slope $\alpha$ of the dark matter density profile ($\rho_{DM} \propto r^{\alpha}$) and the core radius $r_{\rm core}$ of the pseudo-isothermal sphere (see Eq. \ref{isotheq}). Examples of the fits are shown in Figure \ref{core_radius_iso}.

\subsection{Inner slopes of dark matter halo profiles}

We estimate the slope $\alpha$ of the dark matter density profile by fitting a power law relation $\rho_{DM} \propto r^{\alpha}$ in the $1-2\% R_{\rm vir}$ interval. This range is well resolved for all of our main halos at z=0 and it is physically meaningful as it shows the relative profile change at a fixed fraction of the halo size. For dwarf galaxies, this is close to the region where observations indicate shallow and core-like profiles in low-mass galaxies \cite[typically measured at a few hundred pc; see][]{Oh11,Oh15,THINGS,Hunt12}. We also show slopes at $0.5-1\% R_{\rm vir}$ for comparison. Example of the fit for $\alpha$ are shown in Figure~\ref{core_radius_iso}.
We have also measured  $\alpha$  in the fitting ranges of $0.3-0.7$kpc and $1-2$kpc. In Appendix \ref{choiceofalpha} we discuss limitations of these alternative fitting ranges and show that general trends of $\alpha$ with halo mass are similar to our default fitting choice.

\begin{figure}
\includegraphics[scale=0.38]{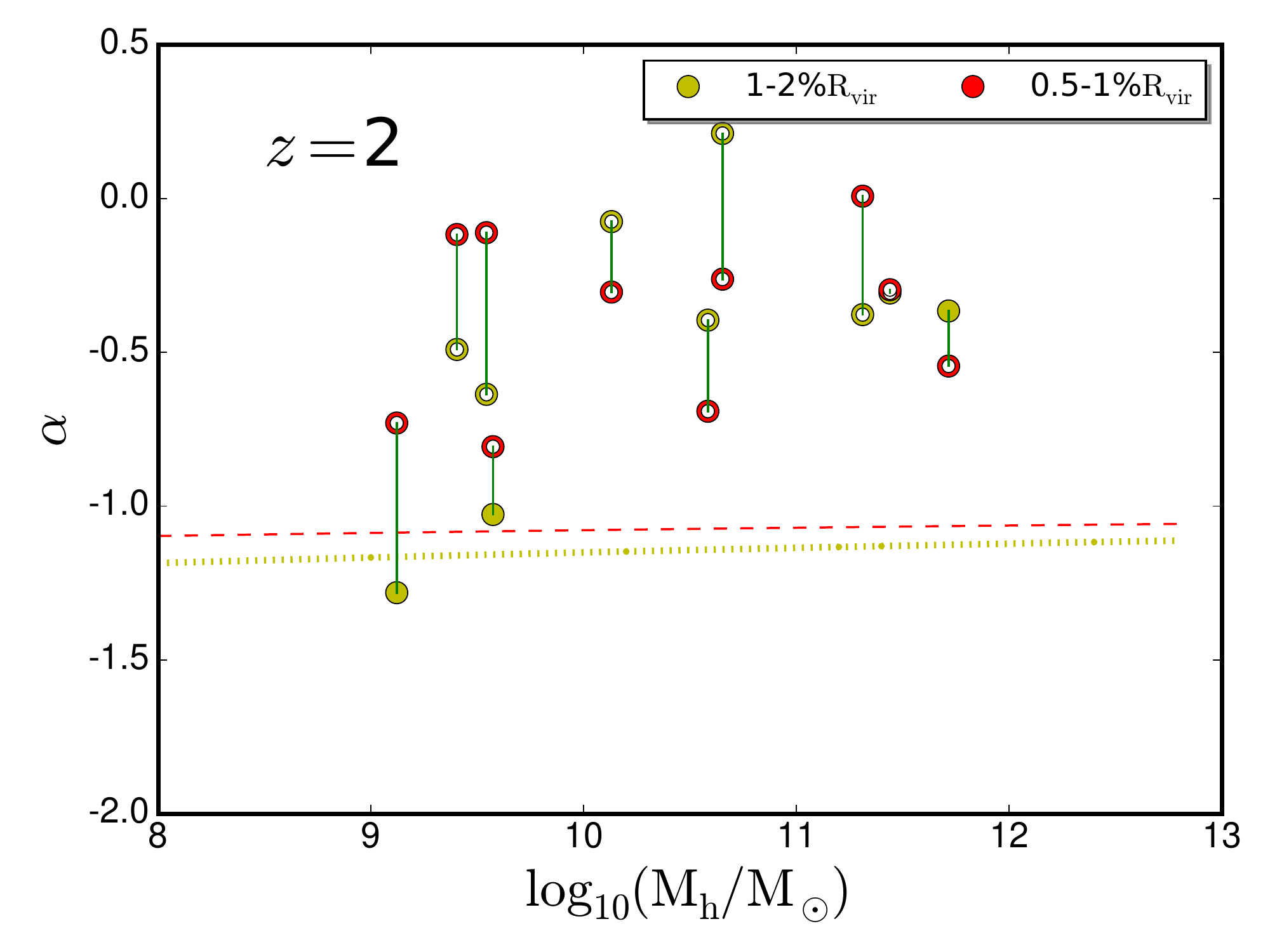}
\includegraphics[scale=0.38]{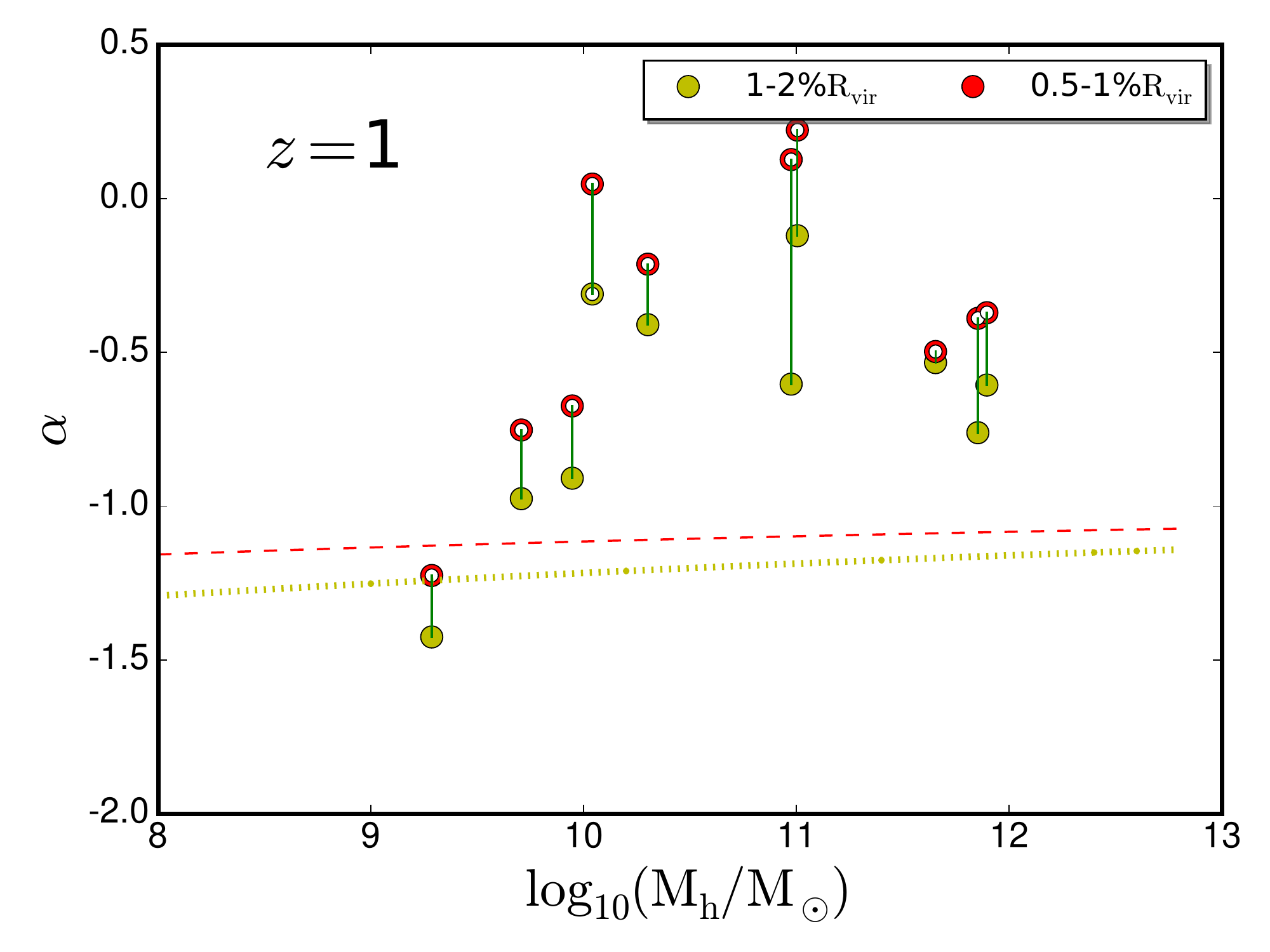}
\includegraphics[scale=0.38]{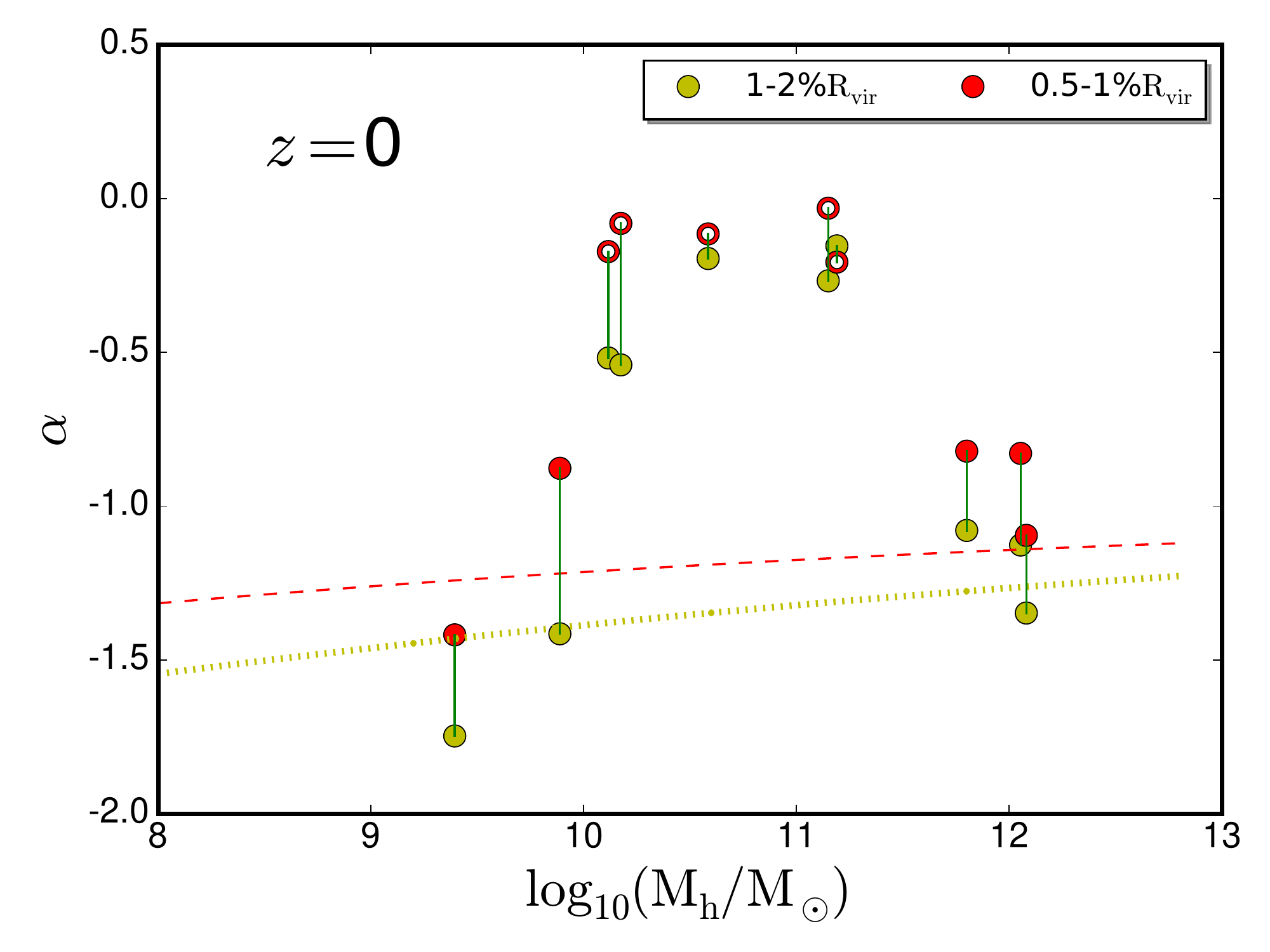}
\caption{Slopes of dark matter density $\alpha$ as a function of halo mass at different redshifts. The green dotted and red dashed lines represent the expected $\alpha$ for the NFW profiles measured at 1-2$\%$  and 0.5-1$\% R_{\rm vir}$ respectively. The concentration of the NFW profiles is evolved with redshift as in \citet{Dutt14}.  Filled circles represent $\alpha$ for simulated halos in which the fitting range is larger than 0.5 $r_{\rm Pow}$ and smaller than one third of $r_{\rm s}$. Hollow circles represent the slopes in halos in which at least one of these criteria is not satisfied (see the main text for other selection criteria) At $M_{\rm h} \sim 10^{10}-10^{11} \msun$, baryonic effects lead to profiles significantly shallower than the corresponding NFW profiles from N-body simulations. }
\label{Mvir_alpha_percent}
\end{figure}

Figure \ref{Mvir_alpha_percent} shows $\alpha$ as a function of the halo virial mass, $M_{\rm h}$, at different redshifts. We only show main halos with more than $10^5$ DM particles and remove all sub-halos and halos with  more than 1\% contamination by more massive DM particles within the inner 0.1$R_{\rm vir}$
\footnote{Note that in most runs we use a ``padding region'' around our zoom in region where mass resolution is lower only by factor of 8. Mild contamination with such particles can sometimes occur within $R_{\rm vir}$ but typically has no consequence on the evolution of halo gas or central slopes.}. The hollow circles represent $\alpha$ whose fitting range contain regions smaller than 0.5 $r_{\rm Pow}$ and/or larger than 1/3 $r_s$, where $r_{\rm s}$ is the scale radius of the NFW profiles. \footnote{We estimate $r_{\rm s}$ from the concentration $c=R_{\rm vir}/r_{\rm s}$ at a given mass, from the concentration-mass relation of \cite{Dutt14}.}

We focus on $z \le 2$ when profiles of halos start to stabilize as rapid halo growth subsides. At $z=0$, the simulated halos show a clear tendency to form shallow central profiles at $M_{\rm h}\sim 10^{10}-10^{11}\msun$. All of the profiles in this range are significantly shallower than the NFW profile. 
More accurate estimate of the halo mass and stellar mass ranges where feedback flattens central slopes will require a larger number of simulations as our statistic are currently limited. When profiles are measured at even smaller radii, 0.5-1\% of $R_{\rm vir}$, profiles are typically even more shallow. At z=2 we see that the scaling with mass shows much larger dispersion, which owes to very bursty star formation and central halo regions that are just coming out of the fast growth stage. We later show that in intermediate mass halos at a fixed physical radius, DM profiles get shallower with time.

\begin{figure}
\includegraphics[scale=0.40]{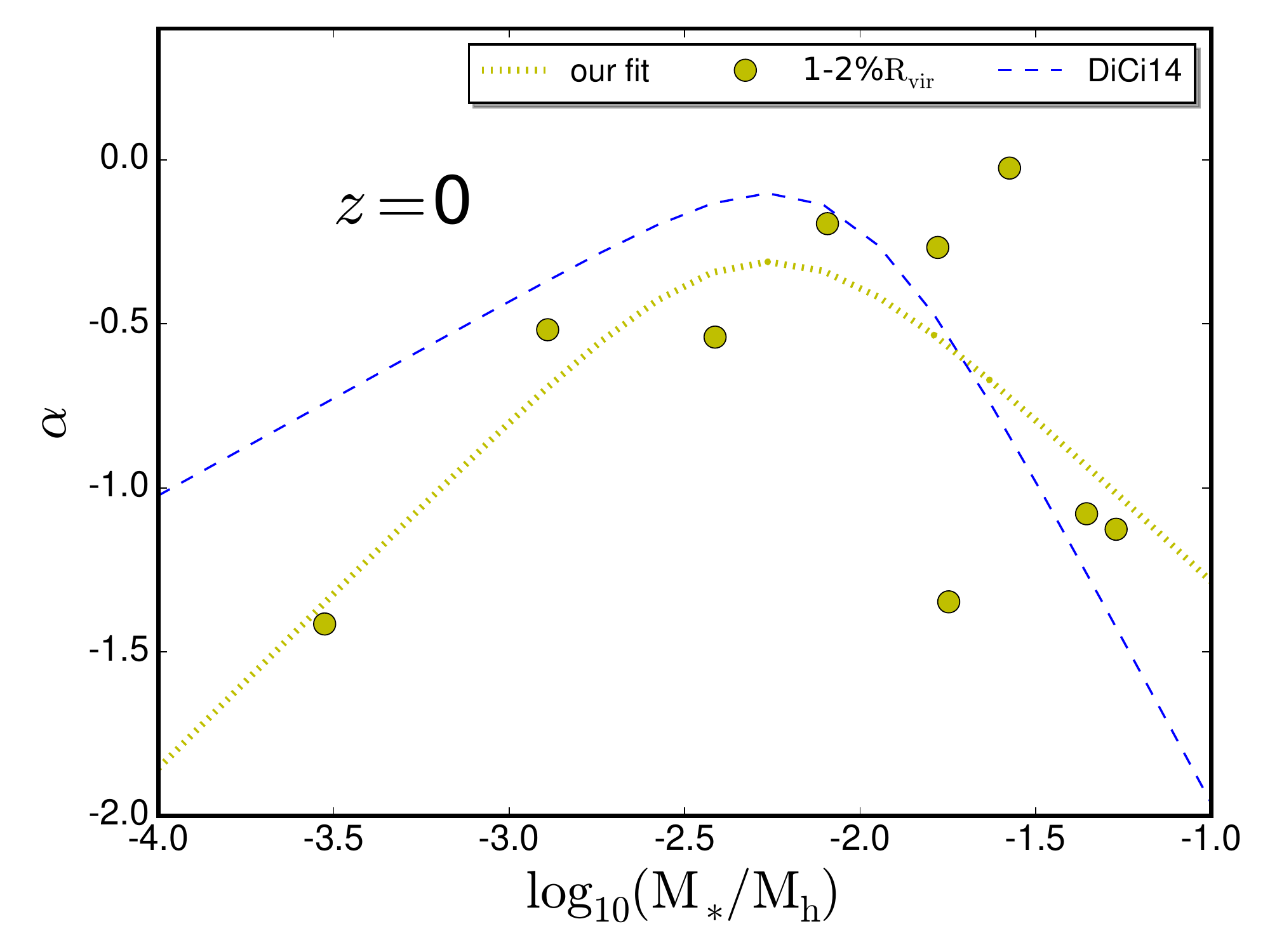}
\includegraphics[scale=0.40]{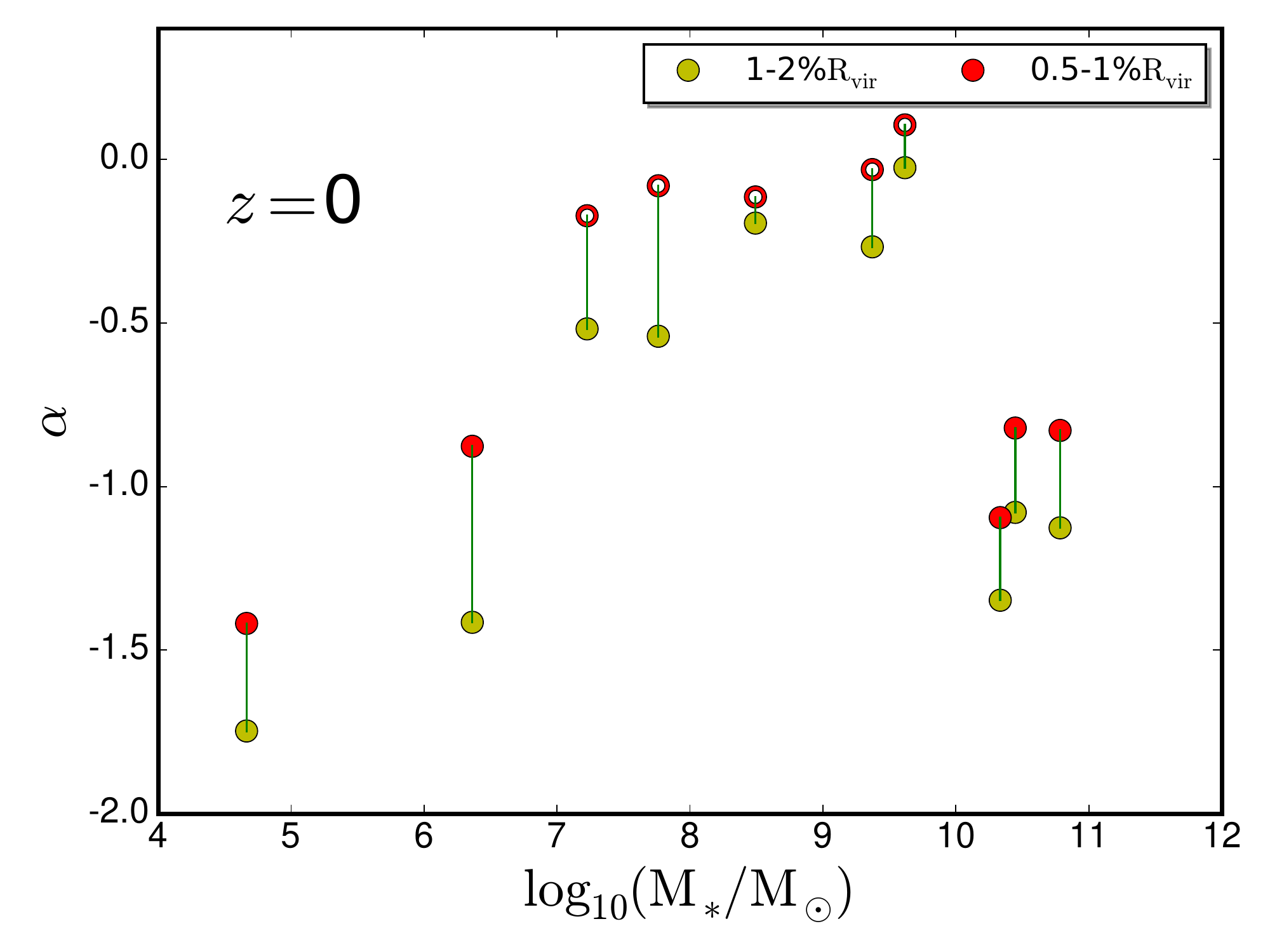}
\caption{({\it Upper}) Relation between $\alpha$ and ratio between $M_*$ and $M_{\rm h}$. The blue dashed line show the fit from \citet{DiCi14} whereas the yellow dotted line shows the same fitting for our data. ({\it Lower}) Relation between $\alpha$ and $M_*$. The symbols are explained in Figure \ref{Mvir_alpha_percent}. The DM profiles near halo centers are cuspy at the lowest and highest masses, and shallowest at $M_* \sim 10^8-10^9 \msun$ and $M_*/M_{\rm h} \sim 0.01$.}
\label{Ms_alpha_percent}
\end{figure}

It is interesting to notice that low mass dwarfs with $M_{\rm h} \ll 10^{10}\msun$ do not develop density cores even at 1\% of $R_{\rm vir}$ (which is typically only several hundreds of parsecs).  As we discuss later, only a small fraction of baryons are converted to stars in these halos, owing to efficient feedback and effects of the UV background. The energy available from a small number of SNe is not sufficient to dramatically modify the dark matter distribution. Around $M_{\rm h}=10^{10}\msun$, the slope of the inner density profile increases rapidly with mass, indicating the development of DM cores. This seems to be a ``threshold'' halo mass needed to develop large cores. As discussed in \cite{Onor15}, small differences in star formation histories in halos close to this threshold can result in the substantial difference in central slopes of the dark matter distribution.

Finally, in halos with mass comparable to the Milky Way ({\bf m12v} and {\bf m12i}) profiles steepen again and are only slightly shallower than NFW. These halos have deep potential wells that can retain a large fraction of available baryons and convert them into stars. Baryons are actually expected to steepen the DM profiles to $\alpha < -1$ owing to adiabatic contraction of dark matter. However, bursty feedback largely cancels and in some cases even overcomes this expected effect of contraction, resulting in slopes $\alpha \gtrsim -1$. The interplay between baryonic contraction and stellar feedback will be discussed in Section \ref{haloexpansion}.

Figure \ref{Ms_alpha_percent} shows the scaling of profile slope with galaxy stellar mass (lower panel) and $M_*/M_{\rm h}$ (upper panel). In terms of stellar mass, feedback significantly modifies DM slopes in the $M_*\sim 10^7-5\times 10^{9}\msun$ range, with a fast transition from cusps to cores occurring at a $\mathrm{few}\times 10^6-10^7 \msun$.  
Overall trends of $\alpha$ with $M_{\rm h}$ in Figure \ref{Mvir_alpha_percent} are similar to the result of \cite{Gove12} and \cite{DiCi14} \citep[see also a recently submitted work by][]{Toll15}. However, we stress that both of these simulations simply suppress cooling in dense gas after supernovae explosions rather than explicitly treat most of the feedback processes around young stars. Furthermore the spatial and mass resolution is typically better in our simulations, by about a factor of ten in mass. This leads to some differences in the slopes of dark matter halos that are illustrated in the upper panel of Figure \ref{Ms_alpha_percent}. In general, profile slopes increase faster with $M_*/M_{\rm h}$ compared to the previous ``subgrid'' models, suggesting faster transition from cusps to cores. We caution that a small number of simulated halos in both samples could also be responsible for some of the differences. In addition we find that the central slope relation is different for the inner 1\% or inner 2\% of $R_{\rm vir}$, which means that the fitting formula in \cite{DiCi14} is not generally applicable. We compare our result with observations in \S \ref{comobs}.

\subsection{Core radii}

In addition to the inner slope, we also examine another parameter, the core radius $r_{\rm core}$ of the halo. We quantify the core size using the the pseudo-isothermal sphere fit that is frequently used to describe dark matter density profiles  (e.g.  \citealt{Bege87,Broe92a,Blok97,Verh97}; another popular fit is the Burkert profile, e.g. \citealt{Salu00}). Density profile is given by
\begin{equation}
\rho(r)=\rho_0\left[1+\left(\frac{r}{r_{\rm core}}\right)^2\right]^{-1},
\label{isotheq}
\end{equation}
where $\rho_0$ is the central dark matter density. We use Eq. \ref{isotheq} to fit the spherically averaged dark matter density profiles of the simulated halos. The two free parameters, $r_{\rm core}$ and $\rho_0$, were determined through a $\chi^2$ minimization fitting procedure starting at $r=0.1\mathrm{kpc}$ and ending at $r=\rm{min}[R_{\rm vir}, 100\rm{kpc}]$.  Table \ref{rcore} lists $r_{\rm core}$ for all halos analyzed in this work. Examples of our fits are shown in Figure \ref{core_radius_iso}. In general fits agree well (to better than few tens of percent within 0.1$R_{\rm vir}$) with the DM density profiles for all halos with $M_{\rm h} < 10^{12} \msun$. For {\bf m12q} and {\bf m12i} pseudo-isothermal profiles deviate significantly from the DM distribution. Evidence of cores is present in all halos. The core size is smallest (relative to $R_{\rm vir}$) in the {\bf m09} run  ($ < 0.5\% R_{\rm vir}$) and largest in {\bf m11h383} and {\bf m10h573} ($> 4\% R_{\rm vir}$) , where we also find the shallowest central slope. Cores of the size of $> 0.005 R_{\rm vir}$ are present even in Milky Way mass halos, albeit with higher central DM density and less shallow central slopes than at $M_{\rm h} \sim 10^{11}\msun$. We compare the core radii from our simulations to  observations in \S~\ref{comobs}.

\begin{table}
	\centering
    \begin{tabular}{lll}
    \hline\hline
    $z=0$     & $r_{\rm core}$ (kpc)& $r_{\rm core}/R_{\rm vir}$  \\ \hline\hline
    {\bf m09}           & 0.17               & 0.0048\\
    {\bf m10}            & 0.38               & 0.0073\\ 
    {\bf m11}          & 4.7                & 0.034\\
    {\bf m12v}           & 1.4                & 0.0061\\
    {\bf m12i}          & 2.0                & 0.0073\\
    {\bf m12q}           & 1.2                & 0.0043\\ \hline \hline
	{\bf m10h1297}               & 2.0                &   0.032     \\
	{\bf m10h1146}              &      2.1              &    0.033     \\
	{\bf m10h573}          &    3.6                &    0.041      \\
	{\bf m11h383}                &    5.7                &     0.041     \\
	\hline    
    \end{tabular}
    \caption{The core radii of the best fitted pseudo-isothermal spheres (Eq. \ref{isotheq}) of the simulated halos at $z=0$.  Large cores of 3-4\% $R_{\rm vir}$ form at $M_{\rm h} \sim 10^{10}-10^{11}\msun$.}
    \label{rcore}
\end{table}

\subsection{Time Evolution of $\alpha$}
\label{timevo}

\begin{figure*}
\includegraphics[width=0.95\textwidth,natwidth=810,natheight=842]{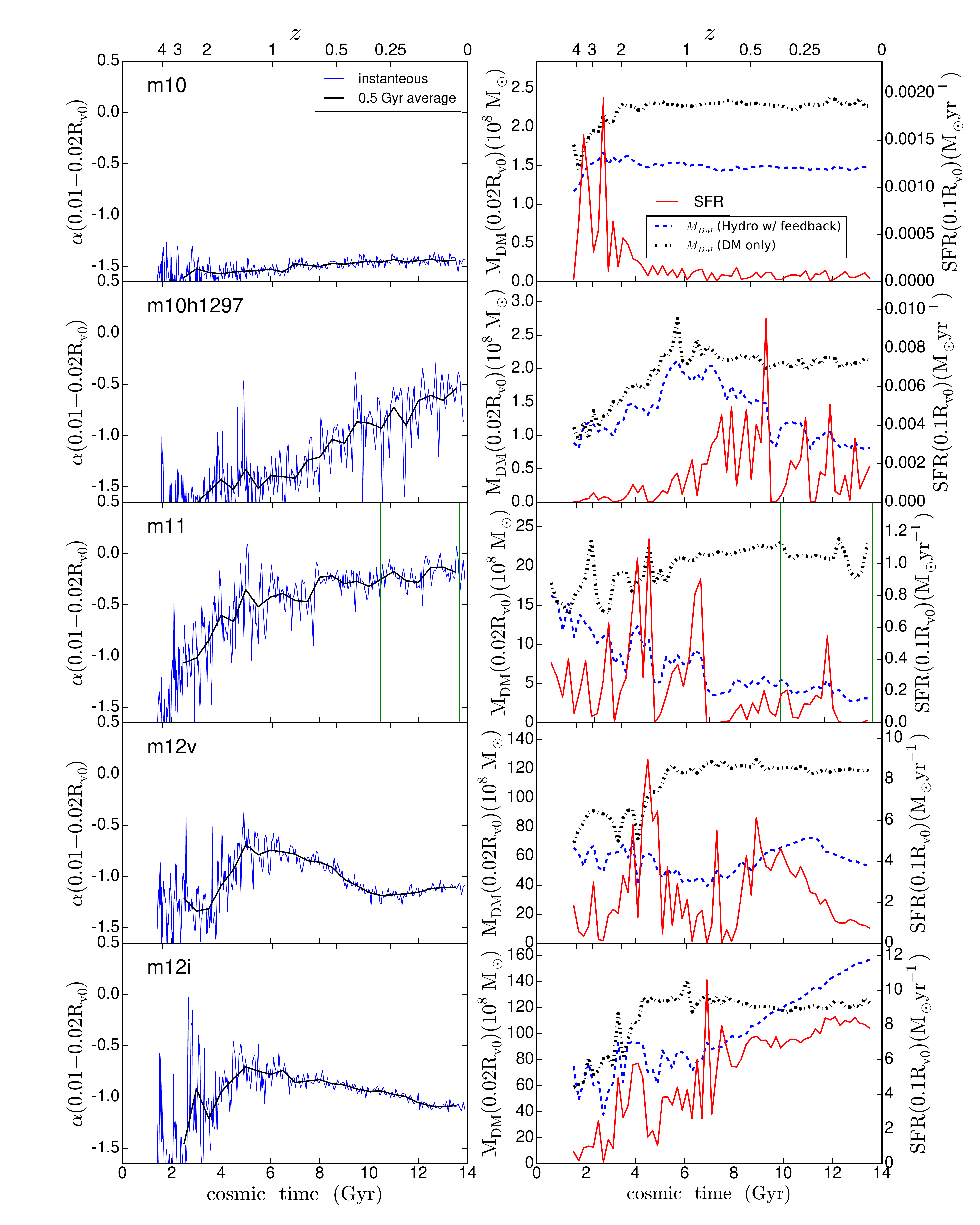}
\caption{ ({\it Left}) Dark matter density slope, $\alpha$, within $1-2\% R_{\rm vir}(z=0)$, $R_{\rm v0}$, as a function of redshift/cosmic time in {\bf m10}, {\bf m10h1297}, {\bf m11}, {\bf m12v} and {\bf m12i} respectively (from top to bottom). The slope is measured at fixed physical radius at all times. Blue lines shows the variations on the time scale of our simulation output (typically ~20-30 Myrs) while the black lines show the average values over 0.5Gyr. ({\it Right}) Time evolution of the enclosed dark matter mass within 0.02$R_{\rm v0}$ for DM only (black dash-dotted) and hydrodynamical simulations with feedback (blue, dashed) and star formation rate within $0.1 R_{\rm v0}$ (red solid), all averaged over 0.2Gyr. Green vertical lines in  {\bf m11} panels show the times of close passages of a subhalo. Cores form when the central DM accretion stops but the star formation is still bursty, as seen in {\bf m10h1297} and {\bf m11}.}
\label{Rviralphaf}
\end{figure*}

Next, we investigate the time evolution of the central slope of the dark matter distribution, for five representative halos from our sample. Left panels of Figure \ref{Rviralphaf} show the time evolution of $\alpha$ measured at $1-2\%$ of the halo virial radius at $z=0$ ($R_{\rm v0}$). For each halo, this radius is kept fixed in physical units at all times. \footnote{We have also analyzed the results within a fixed fraction of the time dependent virial radius ($R_{\rm vir}$ instead of $R_{\rm v0}$) but the correlation between stellar feedback and the enclosed DM mass is difficult to interpret because enclosed DM mass increases with $R_{\rm vir}$.} 
In {\bf m10}, $\alpha < -1$ at all times, which is consistent with its relatively small core size, below $1\% R_{\rm vir}$. In {\bf m10h1297}, $\alpha$ is steadily rising from $-1$ at $z \gtrsim 1$ to $\sim -0.5$ at $z=0$. In {\bf m11}, $\alpha \sim -1$ early on, increasing to $\alpha \sim 0$ around $z\sim 1$ and stays quite flat until late times.\footnote{At late times $z < 0.5$, this halo undergoes several episodes of very fast central slope variations. We examined this system closely and found that these are caused by close passages of a substructure in an ongoing merger. Vertical lines in Figure \ref{Rviralphaf} indicate the times of closest passages: they correlate well with temporary drop and strong oscillations in central slope. The close passages can affect the accuracy of locating center of the galaxy in AHF, which further motivates our two-step center finding procedure described in \S \ref{AHFhalo}.} In {\bf m12i} and {\bf m12v} central DM profiles are flattened to $\alpha \sim -0.7$ at $z\simeq 1$ but steepen afterwards such that central dark matter slope is $\alpha \sim -1$ at the present time.

The right panels show the time evolution of the enclosed mass within $0.02 R_{\rm v0}$ and the star formation rate within $0.1 R_{\rm v0}$ averaged over 0.2 Gyr.
\footnote{Newly formed stellar particles can move quickly between simulation outputs during the star formation burst, owing to feedback induced mass redistribution, so we integrate star formation within this larger radius.} 
\footnote{Results averaged over 0.1Gyr or shorter time-scales are qualitatively similar but show more rapid fluctuations in sub-components of longer bursts so we selected 0.2Gyr for the sake of clarity.}
A dense central concentration of dark matter builds up early in {\bf m10}, with some fluctuations during the bursty star formation epoch at $2 < z <4$, but as star formation subsides the amount of dark matter in the central region remains almost unchanged until the present time. The correlation between star formation and strong outflows of gas that follow the burst \citep{Mura15} and the decrease of dark matter in the central region is clearly visible in the {\bf m10h1297} and {\bf m11} panels. Removal of DM mass occurs after strong bursts of star formation. We examine this more closely in \S \ref{cwsf}.
While some of the dark matter gets re-accreted, the central concentration of dark matter remains lower after the burst for at least several Gyrs.

Each strong burst of star formation reduces central density of DM, so the final slope and core size are a consequence of several burst episodes over a Hubble time. The overall effect is small in {\bf m10}, because the star formation rate decreases to very low values at $z <4$. From the comparison of DM only and feedback simulations it is clear that a small difference in central DM concentration was established early on and stays largely unchanged until late times.

 A small, fluctuating and early decrease in the DM concentration is also seen in more massive halos, e.g. {\bf m10h1297} and {\bf m11}. However, the amount of dark matter in a central region only decreases significantly once the central region finishes rapid growth, at $z \sim 1$ in {\bf m10h1297}  and $z \sim 3$ in {\bf m11}. After this stage, DM only simulations show an approximately constant amount of dark matter in the central region while baryonic simulations successfully evacuate a large amount of dark matter from the center. Unlike {\bf m10}, they have several ongoing bursts of star formation after the rapid-buildup stage, each of which removes a significant amount of dark matter. It appears that having strong bursts of star formation and outflows {\it after} the inner halo buildup slows down is the key to produce a long lasting shallow DM density profile. At early times, during the fast buildup stage, shallow profiles are not fully established as showed in Figure  \ref{Rviralphaf}.

In {\bf m12i} and {\bf m12v}, removal of DM after the peaks of star formation is also seen at $z > 1$ when fluctuations are large. After $z \sim 1$, the star formation continues at modest level without rapid bursts. At the same time, the enclosed DM mass grows slowly. In these massive halos re-accretion of dark
matter in the center occurs when the star formation rate is low and hierarchical assembly is slow. To explain this we followed the central accumulation of baryonic material and found out that in both halos baryons start to dominate central mass at $z \sim 1.5$. As a consequence dark matter gets contracted, increasing the amount in the inner halo (see \S~\ref{haloexpansion_tully}). This effect is stronger in the {\bf m12i} simulation that is more baryon dominated, and accumulates baryons faster at late times. \footnote{In {\bf m12v} the star formation rate is low and the amount of baryons in the central region changes very slowly in the final several Gyrs. The enclosed dark matter at those times is also affected by ongoing minor mergers, that cross very close to the center and are later disrupted causing variations in the enclosed density.}

The core forms after multiple starbursts rather than one single blow-out, which is consistent with the mechanism discussed in \cite{Pont12} \citep[see also][]{Ogiy14}. After a blow-out some re-accretion of DM does occur but even after several Gyrs the amount of DM enclosed in central region does not return to the pre-burst level, indicating that the effect on DM is long lived. These trends are the most obvious at $M_{\rm h} \sim 10^{11} \msun$, i.e. in {\bf m11}. However,  in more massive systems we see that cuspier profiles are re-built at later times. While star formation proceeds to late times, it is often spread throughout the disk. In a companion paper \citep{Mura15} we show that at late times the star formation activity is not able to eject large quantities of material from galaxies, which is why this continuous star formation does not ``heat'' and remove DM from the center. Even if the assembly of central regions of the dark matter halos slows down at relatively early times, increase in central concentration of baryons at late times can rebuilt DM cusps via adiabatic contraction. Our simulations suggest that the contribution of minor mergers to the re-growth of cusps \citep[e.g.][]{Deke03a, Lapo14} at late times is likely sub-dominant or negligible, but a larger number of halos is needed to confirm this for halos with diverse growth histories.

{We have already seen that core formation depends strongly on halo mass but also on the presence of significant star formation episodes after the central region stops accreting dark matter. This is critical at halo masses where large cores start to develop. We showed this directly by using the initial condition of {\bf m10}, simulated with a slightly different local feedback coupling scheme that increases late time star formation and its burstiness: the outcome is a larger DM density core and a shallower center slope compared to the original FIRE {\bf m10} simulation analyzed here \cite[see][]{Onor15} This means that if there are bursts of star formation activity occurring in dwarf galaxies/halos around this mass at late times they could result in shallow density profiles by present time.

The star formation history of {\bf m10} is significantly different from {\bf m10h1297}. It forms most of the stars before $z=2$ and becomes passive at late time whereas other halos remain actively star forming at present. From the observations of dwarf galaxies in the Local Group, \cite{Weis14} show that a large fraction of dwarf galaxies have active late time star formations, especially for field galaxies and galaxies with $M_*>10^6 \msun$. Therefore, the star formation histories of our simulated galaxies are well within the observed range. Our limited statistics suggest that at $M_* \sim 10^6-10^7 \msun$ SFR history is closely linked to formations of large cores \citep[see also][]{Onor15}.

\subsection{Halo Expansion or Baryonic Contraction?}
\label{haloexpansion}
In the previous section, we investigated the shapes of profiles of our simulated halos. Here we examine the effect of a central concentration of baryons on the dark matter profiles.

First, we examine the net effect of halo expansion via feedback and halo contraction owing to central concentration of baryons. Feedback is dominant in shaping flatter profiles in lower mass halos, but baryonic contraction largely cancels the feedback effect in  Milky Way mass halos, such that their final profiles are only slightly shallower than the NFW profile.

In order to estimate the effect of baryonic contraction on the dark matter profiles, we follow \cite{Blum86} to calculate the final dark matter mass distribution $M_x$, given the final baryon mass distribution $m_{\rm b}$ and the initial total mass distribution $M_i$:
\begin{equation}
r[m_{\rm b}(r)+M_x(r)]=r_iM_i(r_i)=r_iM_x(r)/(1-F_b),
\label{barconeq}
\end{equation} 
where $F_b$ is the fraction of dissipational baryons, $r_i$ and $r$ are initial and final orbital radii respectively. 
We follow the simple semi-analytic model \citep{Dutt07}, and assume that initial total halo density profile is the one from our DM only simulation, and use the final distribution of baryons (stars and gas) in the full physics simulation to estimate the contraction. Therefore, we use $m_{\rm b}$ from hydrodynamical simulations with stellar feedback, and $M_i$ is the halo mass from dark matter only simulations. 
\footnote{We also tried to account for the loss of baryons due to feedback but the mass loss is much smaller than the total mass, especially in {\bf m12} series. The difference in circular velocity with or without the missing baryons is less than a few percent in {\bf m11} and negligible at higher masses.} 

Figure \ref{densityf} shows density profiles from collisionless simulations, full feedback simulations, and from models with a contracted DM halo.  Left panels show three different DM profiles for each halo and baryon density profiles from feedback simulations for reference. Right panels show the total density profiles including DM and baryons (gas and stars).

In {\bf m10}, the estimated effect of adiabatic contraction is very small because the fraction of baryons in the center of {\bf m10} is small. The baryon density is less than one tenth of dark matter density near the center. However, the feedback slightly expands the DM and forms a small core.

In {\bf m10h1297} and {\bf m11}, stellar feedback strongly affects dark matter distribution, making it much shallower than in the corresponding collisionless run. It is interesting that in {\bf m10h1297}, feedback significantly flattens the DM profile, creating a large core, but the halo remains dark matter dominated at all radii. In {\bf m11}, the baryon density is a slightly larger fraction of the total, but still significantly lower than the corresponding dark matter density in the collisionless simulation. The contracted profiles for both of these halos are therefore very similar to the dark matter profiles from the corresponding collisionless simulations.

In more massive halos, the {\bf m12} series, baryonic contraction is expected to significantly steepen the DM profiles, because their central regions are baryon dominated. While {\bf m12v} shows a strong effect of feedback, the profile of {\bf m12i} is relatively steep in the center, similar to the NFW profile. However, when compared to the expectations of baryonic contraction we see that the resulting profile is much shallower than contracted NFW halo for all of the plotted simulations. This demonstrates that even in these massive halos, feedback has a strong effect on shaping the final dark matter profiles, and largely cancels out the effect of contraction. In general the expansion of the halo by the stellar feedback causes an order of magnitude difference in the density profiles around 1 kpc in $10^{11-12}\msun$ halos when compared to expectations of a simple baryonic contraction model. 

While feedback effects on the DM distribution are substantial even in {\bf m12} series halos, in those halos baryons completely dominate the central few kpc at z=0. This is why the differences between the total matter density profiles in simulations with feedback and profiles expected from the contracted original DM halo are relatively small (right panels). We will return to these total matter profiles shortly and show that even this small effect has measurable consequences in the circular velocity curves of galaxies. It is interesting to note that the total matter distribution in the inner 20\% of the $R_{\rm vir}$ of {\bf m12} series is well approximated by the isothermal density profile ($\rho \propto 1/r^2$).

The results from our {\bf m12v} simulation are consistent with the strong feedback run in \cite{Macc12b} who showed core formation in a $7\times10^{11}\msun$ halo. We do however find slightly higher central density of dark matter than reported in \cite{Macc12b}. While this might be just a matter of small number statistics, our simulation results should give more accurate predictions for the central profiles both because of more realistically implemented stellar feedback and because of the higher resolution. At masses similar to the Milky Way ($M_{\rm h} \sim 10^{12}\msun$), the dark matter density distribution in the center is only slightly shallower that the NFW profile as strong feedback effects and adiabatic contraction of such expanded halo nearly cancel out.

\begin{figure*}
\includegraphics[width=0.95\textwidth,natwidth=810,natheight=842]{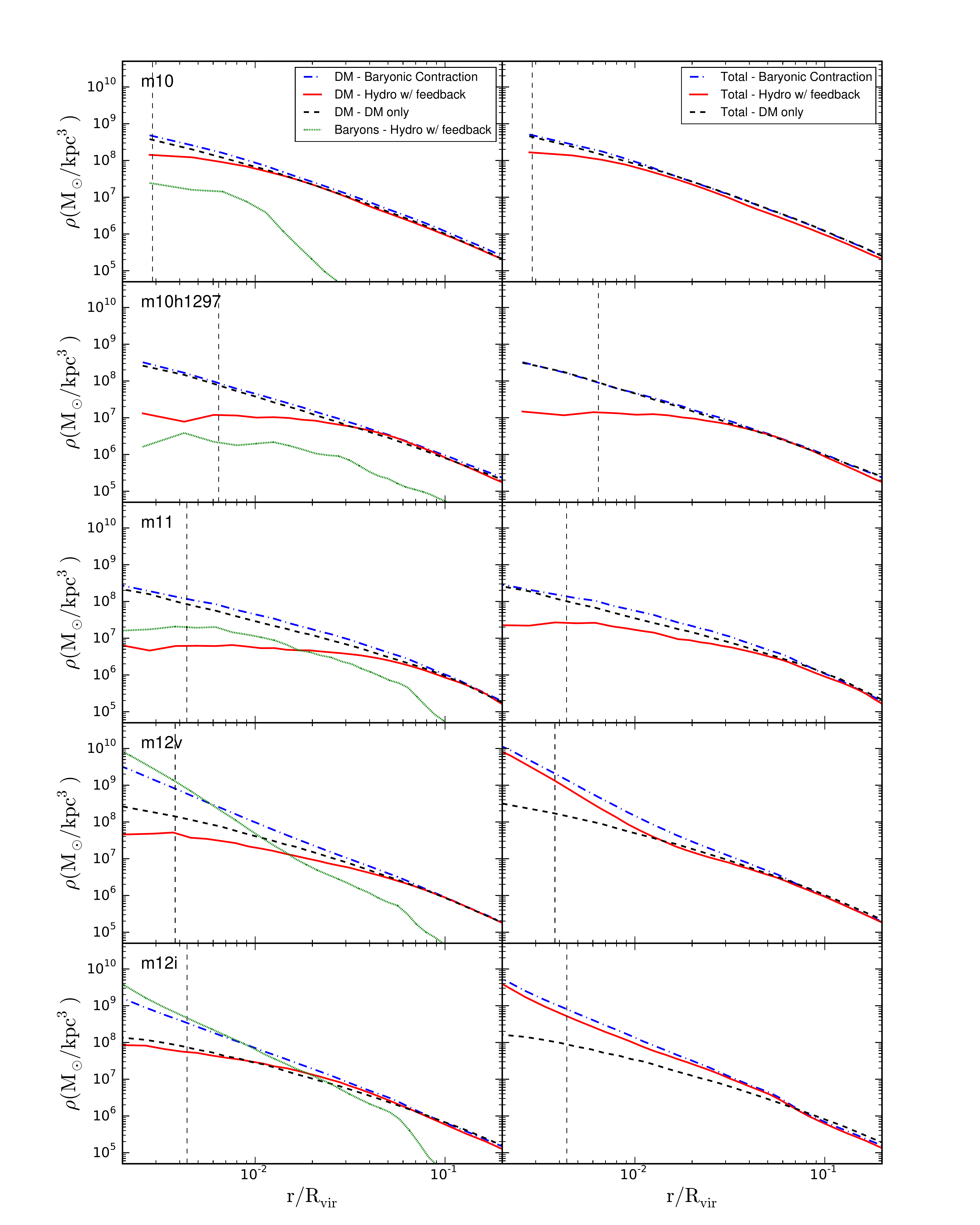}
\caption{({\it Left}) Dark matter density profiles of {\bf m10}, {\bf m10h1297}, {\bf m11}, {\bf m12v} and {\bf m12i} at $z=0$. Different line colors show the expected DM profile in simulations with baryons and feedback (solid; red), in collisionless simulations (dashed; black) and in calculations including baryonic contraction (dot-dashed; blue). The green dotted line shows the total baryon density, including both gas and stars, in feedback runs. ({\it Right}) Total density profiles of the same halos, including both dark matter and baryons. The Power convergence radii are shown as dashed vertical lines. In halos where baryons dominate in the central regions, total and dark matter densities based on the simple baryonic contraction model are higher than the actual densities in our simulations with baryonic feedback. Feedback effectively cancels the effect of baryonic contraction.}
\label{densityf}
\end{figure*}

\section{Observable consequences}

\label{haloexpansion_tully}

The following subsections show how our simulations with stellar feedback can alleviate the tensions between previous simulations and observations, including the ``lack'' of baryonic contraction, the ``Too Big To Fail'' and {\it cusp/core} problems.

\subsection{Rotation curves and the Tully-Fisher normalization}

The distribution of matter in galaxies can be measured with the rotation curves. For disk galaxies, there is a tight relation between their luminosity (or mass) and their circular velocity, so called the Tully-Fisher \citep{Tull76} relation (TFR). Here we examine the effect of stellar feedback and baryonic contraction on rotation curves.

Figure \ref{vrot} shows the rotation curves of halos from simulations and baryonic contraction calculations for {\bf m11}, {\bf m12v} and {\bf m12i}. Here $V_c=\sqrt{GM(r)/r}$. The rotation curves of simulated galaxies do not show profiles expected from the NFW halos affected by adiabatic contraction. Instead, galaxies have lower mass concentrations in the centers, resulting in lower circular velocities. Therefore feedback effectively prevents buildup of high densities expected from strong adiabatic contraction. As expected, the rotation curves in the {\bf m12} series are approximately flat at radii larger than several kpc, while for the massive dwarf {\bf m11} the rotation curve is rising.

The effect of feedback, which cancels the effect of baryonic contraction, turns out to be very important for the normalization of the TFR. We do not perform a detailed comparison with observations, which would require mimicking observational measurements of the rotational velocity and luminosity. We plan to study this in future work. Here, instead, we focus on the relative effect of the feedback on dark matter profiles that determine the normalization of TFR. To show this effect and compare our simulated galaxies with the observed TFR, we measure the circular velocity of the halo $V_c=\sqrt{GM(r)/r}$. We measure circular velocity at 2.2 `disk scale length' $V_{2.2}$, approximately mimicking frequent observational approach \citep{Dutt10}. We first measure the half mass radius of the stellar distribution $r_{1/2}$  and use the relation between half mass radius and scale radius of an exponential disk to define disk scale length as $r_d = r_{1/2}/1.67$.

Figure \ref{TullyFisher} shows the TFR of main galaxies with $10^9<M_*/M_\odot<10^{11}$, and the best fit of the observed TFR from \citet{Dutt10}.  They derived the TFR using the data from \cite{Cour07} and with the best fit:

\begin{equation}
\log_{10}\frac{V_{2.2}}{[\mathrm{km}\, \mathrm{s}^{-1}]}=2.064+0.259\left(\log_{10}\frac{M_{*}}{[10^{10}\msun]}\right).
\label{Duttoneq}
\end{equation}

This relation was derived for relatively massive galaxies, most with ${\rm log V_c [km/s] > 1.8}$. Similar relations were found for dwarfs, however with significantly enlarged scatter and non-uniform way of measuring $V_c$ \citep[e.g.][]{Ferr12}. We therefore limit discussion here to galaxies with $M_*>10^9 \msun$.

\begin{figure}
\includegraphics[scale=0.43]{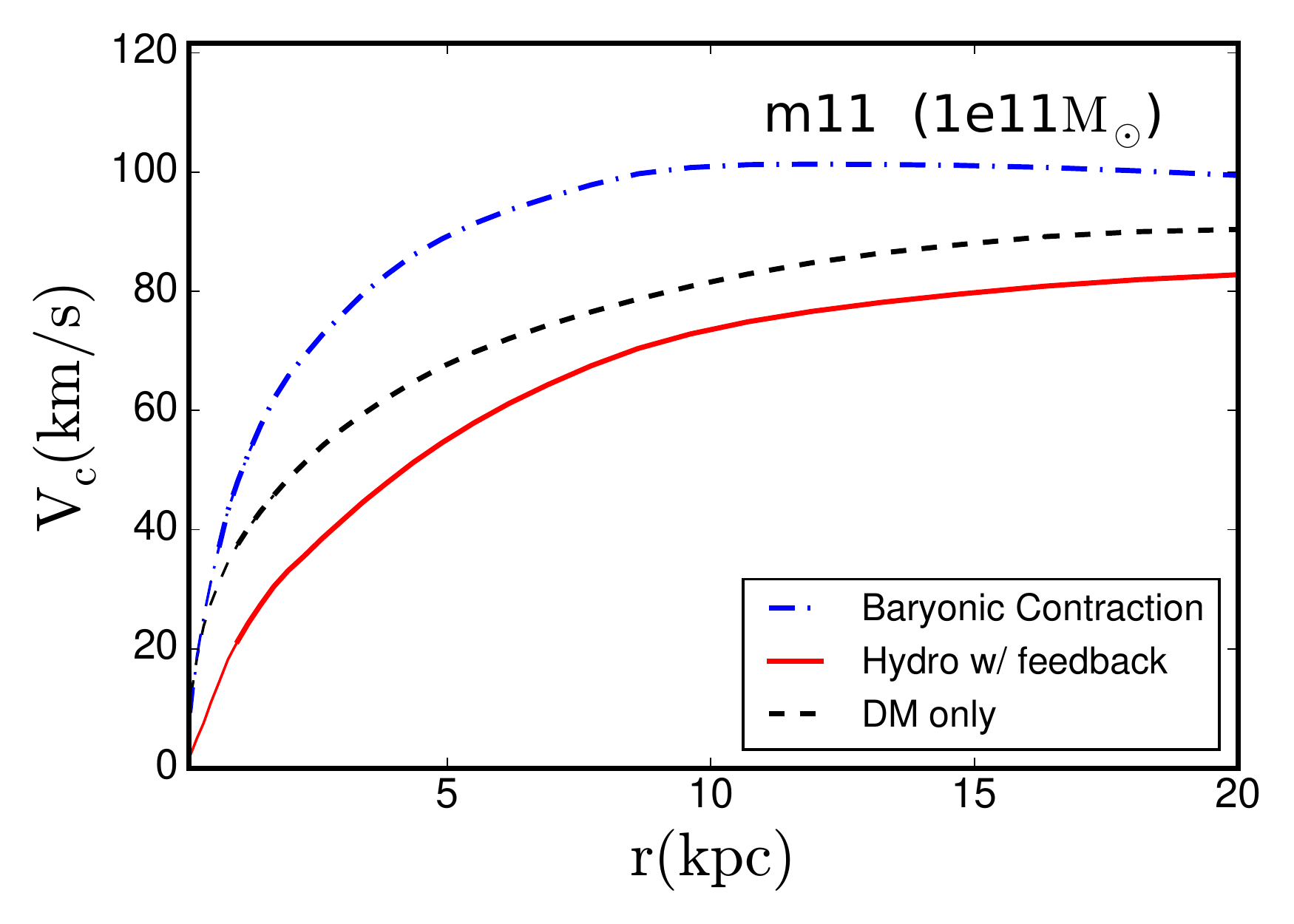}
\includegraphics[scale=0.43]{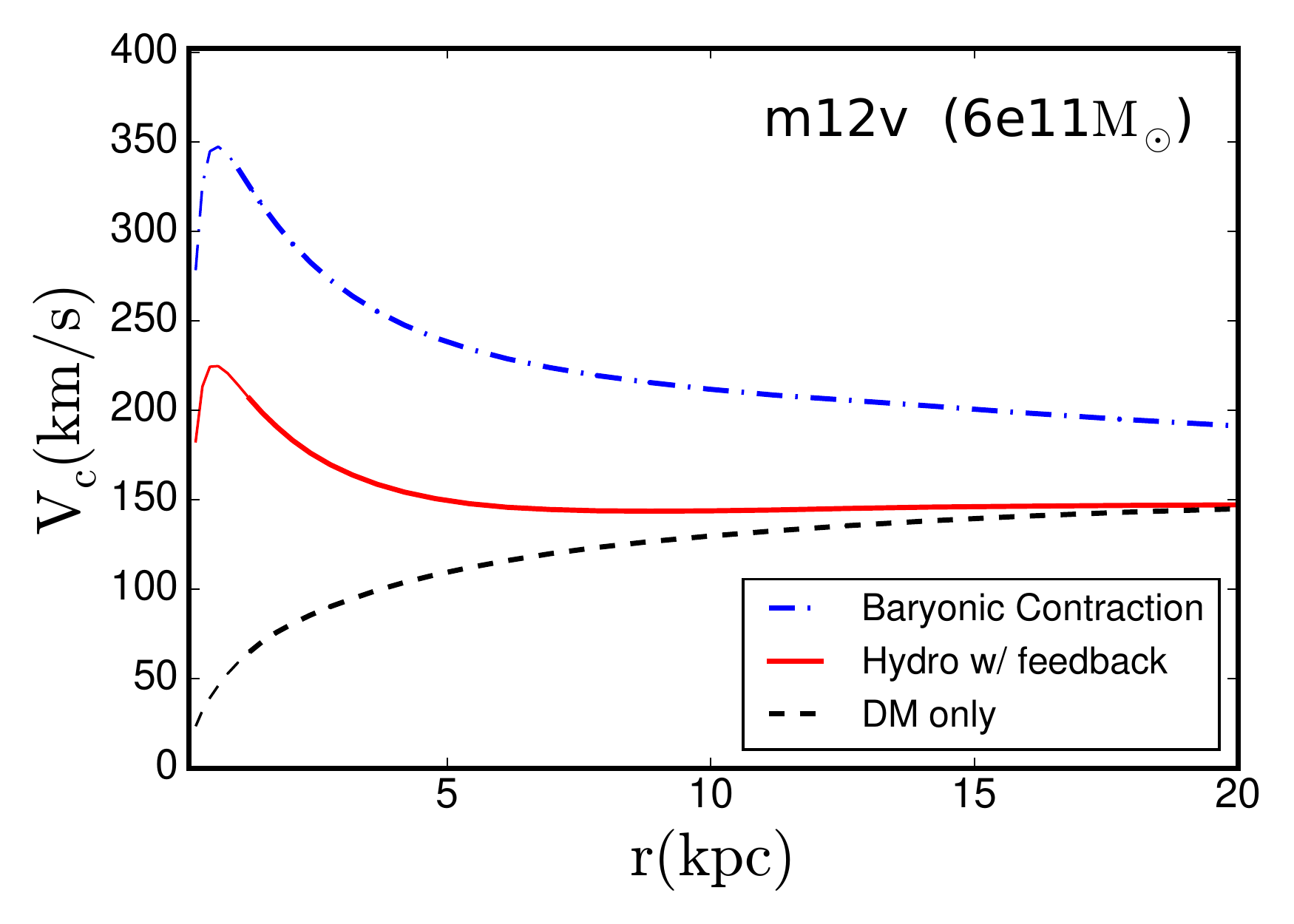}
\includegraphics[scale=0.43]{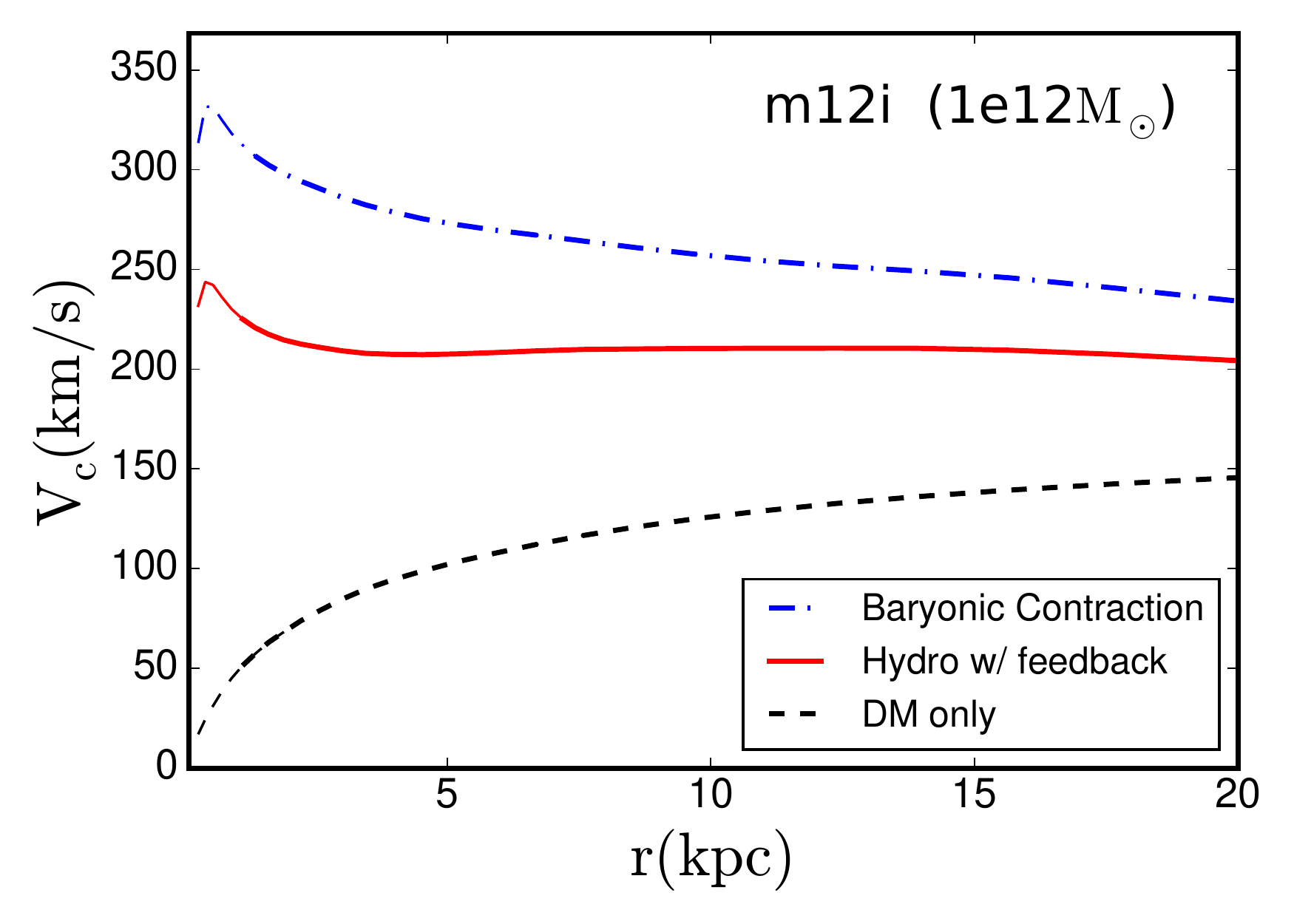}
\caption{Rotation curves of halos, including {\bf m11}, {\bf m12v} and {\bf m12i}. Red solid lines represent rotation curves from simulation with feedback, black dashed lines show results from DM only simulations, whereas blue dot-dashed lines represent rotation curves with baryonic contraction. Dashed lines show the region within the Power radius. Their halo masses are shown in the brackets. The rotation curves in simulations with baryonic feedback are lower than in a simple baryonic contraction model. The simulated Milky Way-mass halos show flat rotation curves, while the large dwarf galaxy {\bf m11} shows a rising rotation curve.} 
\label{vrot}
\end{figure}

It is clear that the strong feedback which reduces the effect of adiabatic contraction is a necessary ingredient in reproducing and explaining the TFR in massive galaxies with $M_*>10^{10}\msun$. While there are direct effects of feedback on the distribution of baryons within galaxies feedback effect on the distribution of dark matter is also an important ingredient in establishing the TFR. Simulated galaxies appear to better match the observed TFR than our model with baryonic contraction. The circular velocities in baryonic contraction calculations are higher by a factor of 1.2-1.5 than $V_c$ in simulations.

Our findings confirm previous conclusions that the lack of effective contraction is necessary to explain the Tully-Fisher relation \citep{Dutt07,Macc12b}. This also explains why previous generations of simulations without efficient feedback had trouble matching the normalization of TFR \citep[e.g.][]{Stei00}. In these models too much gas collapsed to the center exerting strong contraction of DM halo without previously affecting the DM distribution.

\begin{figure}
\includegraphics[scale=0.45]{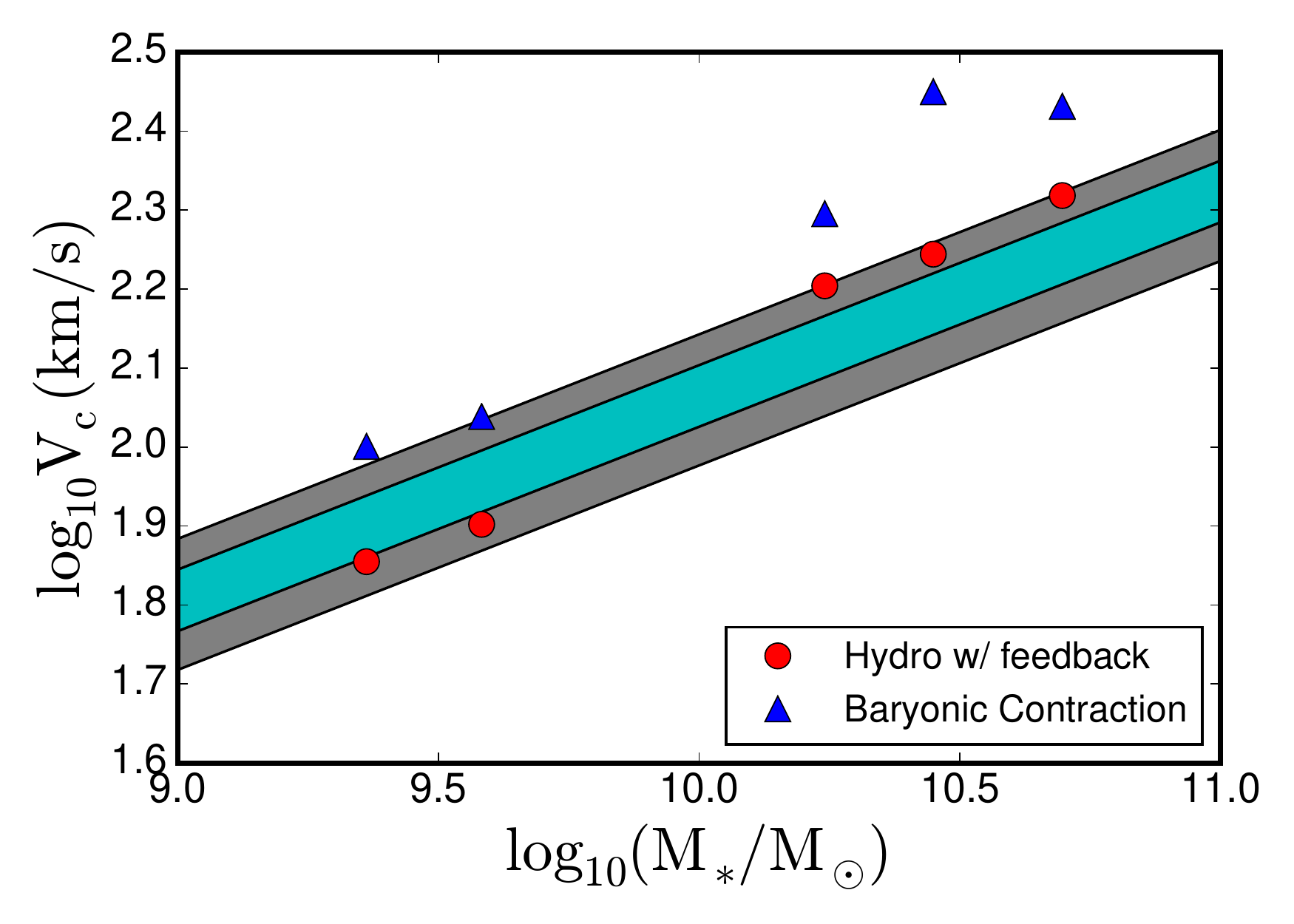}
\caption{ Stellar mass --- Tully-Fisher relation for observed galaxies (Eq. \ref{Duttoneq}) and for the simulated halos ({\bf m11h383}, {\bf m11}, {\bf m12v}, {\bf m12q} and {\bf m12i}). The y-axis is the rotation velocity measured at 2.2 disk scale length. The shaded regions show Eq. \ref{Duttoneq} with one sigma uncertainty ($\sigma=0.039$)(cyan) and two sigma uncertainty on the zero-point (grey) respectively. Stellar feedback, which counteracts the effects of adiabatic contraction, appears necessary to establish the observed normalization of the Tully-Fisher relation.}
\label{TullyFisher}
\end{figure}

\subsection{Implication for the `Too Big To Fail' problem}

In addition to the cusp/core problem, cold dark matter simulations are also challenged by another problem, the so called ``Too Big to Fail'' problem \citep{Boyl11,Garr13}. The simplest version of this problem is that the observed Milky Way's satellite galaxies have much lower central circular velocities than sub-halos from cosmological collisionless cold dark matter simulations. This either means that massive sub-halos do not have corresponding match in observed satellite galaxies or that central regions of predicted cold dark matter sub-halos are too dense compared to observed halos. This seems to be generic problem, independent of the halo formation history as similar effects are also observed in the Local Group and for dwarf galaxies in general \citep{Ferr12,Garr14,Papa15} including non-satellite galaxies. 

We have already shown that stellar feedback can reduce the central dark matter density, which could potentially resolve this discrepancy, without invoking different a type of dark matter.
In Figure \ref{tbtf}, we show the circular velocity profiles of the central kpc of {\bf m10}, {\bf m10h1297} and {\bf m10h1146} at z=0 along with their corresponding dark matter only simulations. These are our best resolved systems with galaxy stellar mass $\sim 2\times 10^6-8\times 10^7\msun$ at z=0, which are close to the stellar masses of the galaxies for which the ``Too Big to Fail'' problem was demonstrated \citep{Boyl11,Garr13}. For comparison we also include the observational data from Milky Way satellite galaxies \citep{Stri07,Walk09,Wolf10} and Local Group field galaxies \citep{Kirb14}. 
It is clear that feedback strongly reduces circular velocities in the central few hundred pc with respect to collisionless cold matter simulations, in some cases by a large factor.  Such reduced circular velocity implies that observed dwarf galaxies (including satellite galaxies) should not be associated with halos and sub-halos from DM only simulations with the same circular velocity, instead these should be connected to the predicted higher circular velocity analogs, whose circular velocity is now reduced owing to feedback. Our results strongly suggest that this effect would dramatically reduce the number of ``massive failures'' and can alleviate or potentially solve the ``Too Big to Fail'' problem. Our findings qualitatively agree with hints in previous work \citep{Broo15}.

It is interesting that the high stellar mass dwarf galaxies (e.g. {\bf m10h1146} and {\bf m10h1297}) have more significant reductions in central rotation velocities (and thus dynamical masses) compared to low stellar mass dwarf galaxies (e.g. {\bf m10}). This causes the rank order of $V_c$ at small radius (e.g. 500pc) not to correspond to the rank order of their $V_{max}$ or their stellar mass, as illustrated in the middle and lower panels of Figure \ref{tbtf}. Direct comparison between dark matter only simulations and simulations with baryons is even more complex, and rank order matching of $V_c$, or $V_{peak}$ from measurements at small radii might lead to incorrect physical interpretations.

Our results only indirectly address the ``Too Big To Fail'' problem in satellite galaxies, because the galaxies we consider here are not satellites but field galaxies.  Satellite galaxies of relevant mass in our {\bf m12} simulations do not have sufficient mass resolution to study their dark matter distributions at the galaxy centers. The effect of host galaxies on satellites, e.g. tidal stripping, could also modify the structure of DM halos \citep{Zolo12, Broo14}. 
However, the effect of lowering the circular velocity of galaxies is generic in the range of stellar masses $2\times 10^6- 3\times 10^9\msun$ and we therefore believe that satellite galaxies in this mass range will be affected in the same systematic way.

\begin{figure}
\includegraphics[scale=0.45]{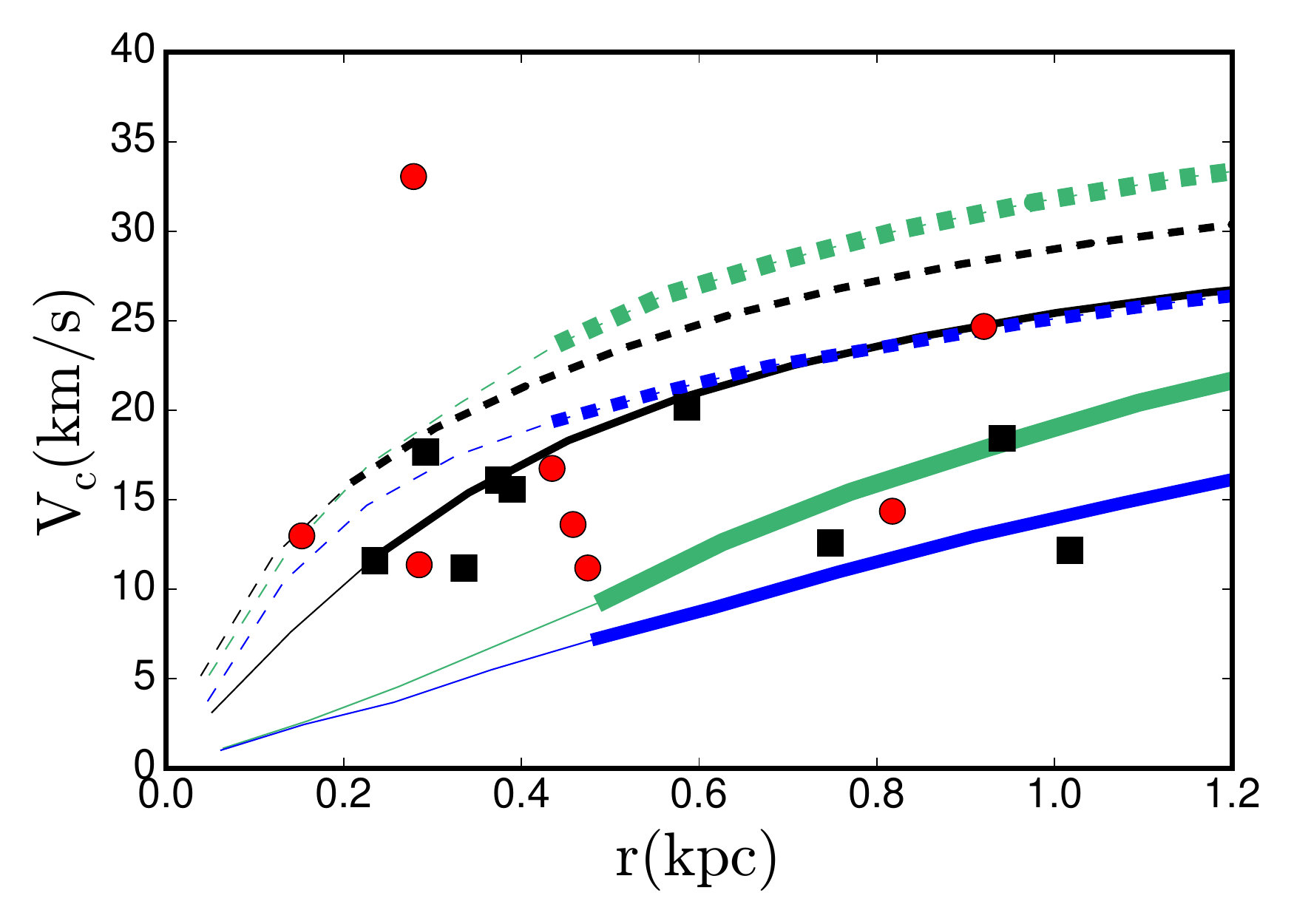}
\includegraphics[scale=0.45]{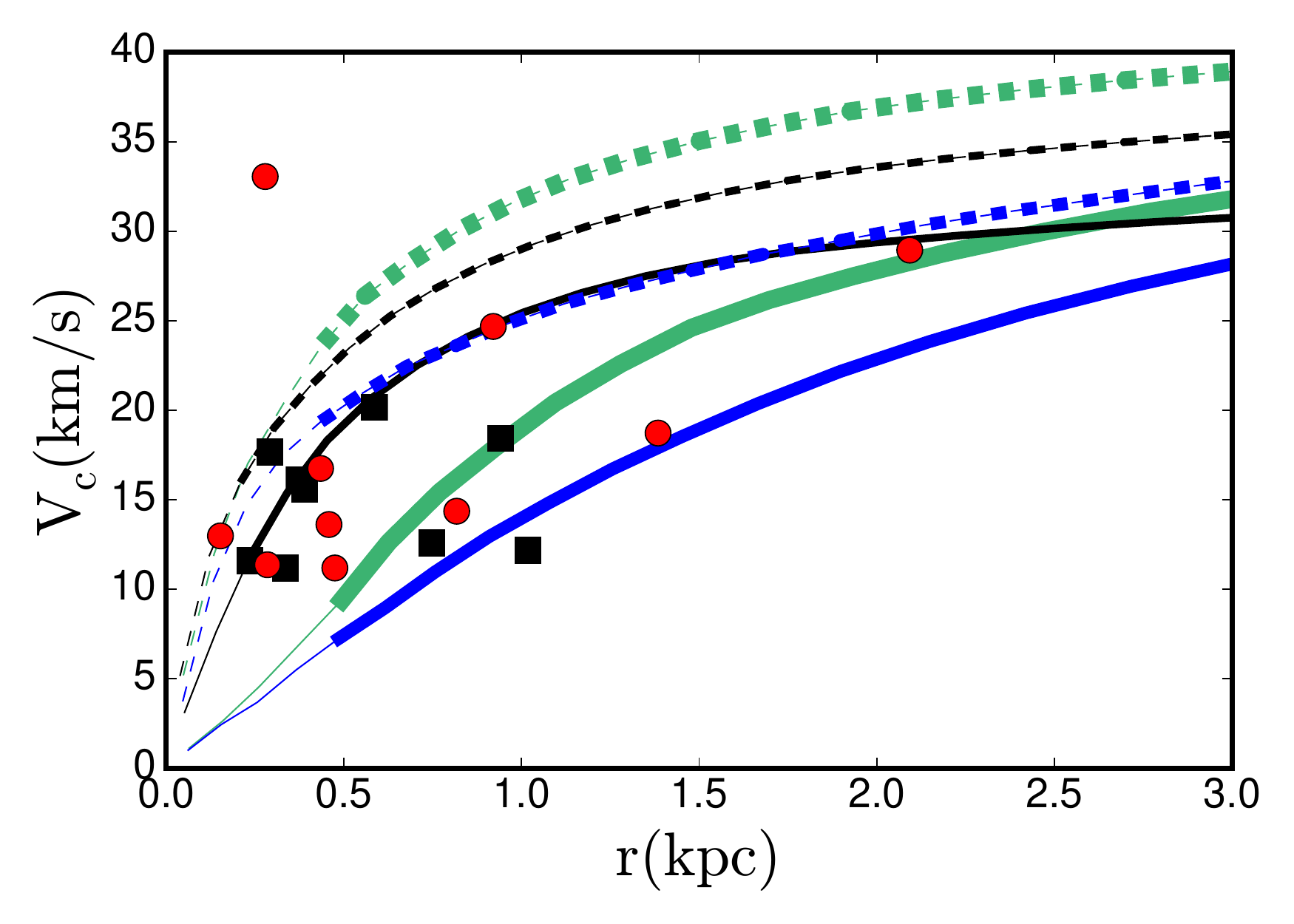}
\includegraphics[scale=0.45]{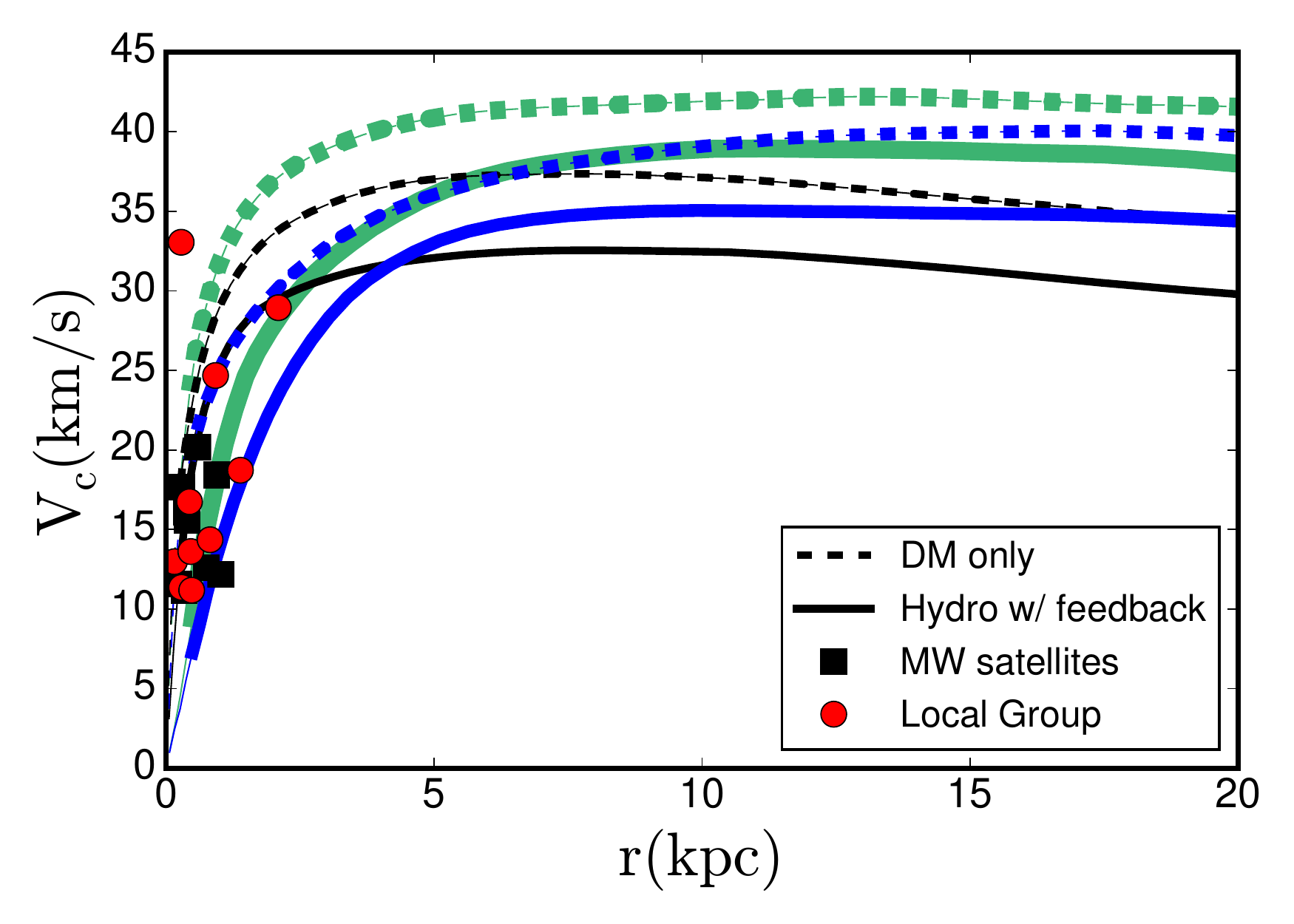}
\caption{Rotation curves illustrating the TBTF problem, plotted over a range of radial scales. We have included halos {\bf m10} ($M_*=2.3\times 10^6M_\odot$; black, thin), {\bf m10h1297} ($M_*=1.7\times 10^7M_\odot$; blue, normal) and {\bf m10h1146} ($M_*=7.9\times 10^7M_\odot$; green, thick) and their corresponding DM only simulations. Thick dashed lines represent the halos from the collisionless simulations while thick solid lines represent the halos from the hydrodynamical simulations with feedback. The thin lines show the velocity curves at radius smaller than the Power convergence radius. Black squares show the data from Milky Way bright satellite galaxies \citep{Stri07,Walk09,Wolf10}, while red circles show the isolated dwarf galaxies in the Local Group \citep{Kirb14}.  The panels show three different scales of the same plot to illustrate that the order of rotation curves at small scales may not imply the order of halo masses. Stellar feedback reduces the central circular velocity such that the rotation curves can match the observed dwarfs, suggesting that baryonic feedback may solve the ``Too Big To Fail'' problem. Observational errors are typically smaller than a few km/s (not shown for clarity).  }
\label{tbtf}
\end{figure}

\subsection{Central Slopes and Core sizes}
\label{comobs}
\begin{figure}
\includegraphics[scale=0.43]{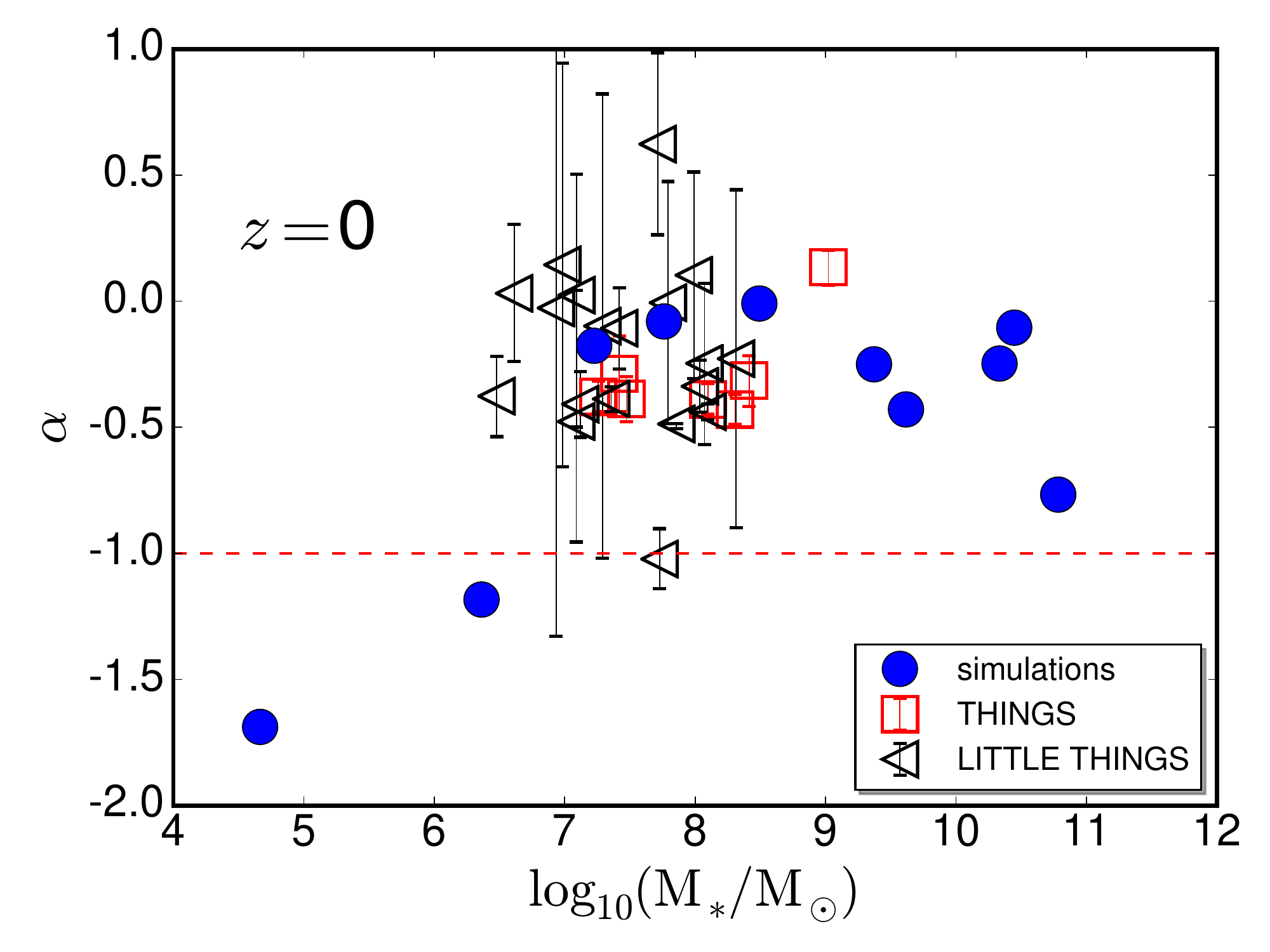}
\caption{Slope of dark matter density profile, $\alpha$, from our simulations (measured at $r=0.3-0.7{\rm kpc}$) compared with $\alpha$ from observations (typically measured at a few hundred pc). Blue solid circles represent the simulated halos at $z=0$. Red hollow squares represent the observed dwarf galaxies from THINGS \citep{Oh11, THINGS}, whereas black hollow triangles represent the dwarf galaxies from LITTLE THINGS \citep{Oh15,Hunt12}. In the overlapping mass range, the simulated dwarf galaxies have central DM profile slopes in good agreement with the observed dwarfs.}
\label{THINGS}
\end{figure}

\cite{Oh11} and \cite{Oh15} measured detailed density profiles of field dwarf galaxies, and found flat DM profiles near the center, in contrast with the cuspy NFW profiles, predicted from N-body simulations. In massive galaxies, such as the Milky Way, baryons dominate in the central few kpc \citep[e.g.][]{Cour15} making measurements of central dark matter properties extremely difficult. We therefore focus on lower mass galaxies/halos with reliably measured central DM profiles.

In Figure \ref{THINGS}, we show that FIRE halos at $z=0$ are in good agreement with the observed slopes of central DM density profiles \cite[see also][]{Gove12}. This suggests that the inclusion of stellar feedback in our simulations helps resolve so called ``cusp/core'' problem observed in low mass galaxies. However, the observed scatter is large and number of simulated objects is limited in the observed mass range. It is therefore clear that much larger sample of model galaxies/halos as well as more detailed accounting for methodology and selection used in observations are needed to test our model in detail.

\begin{figure}
\includegraphics[scale=0.43]{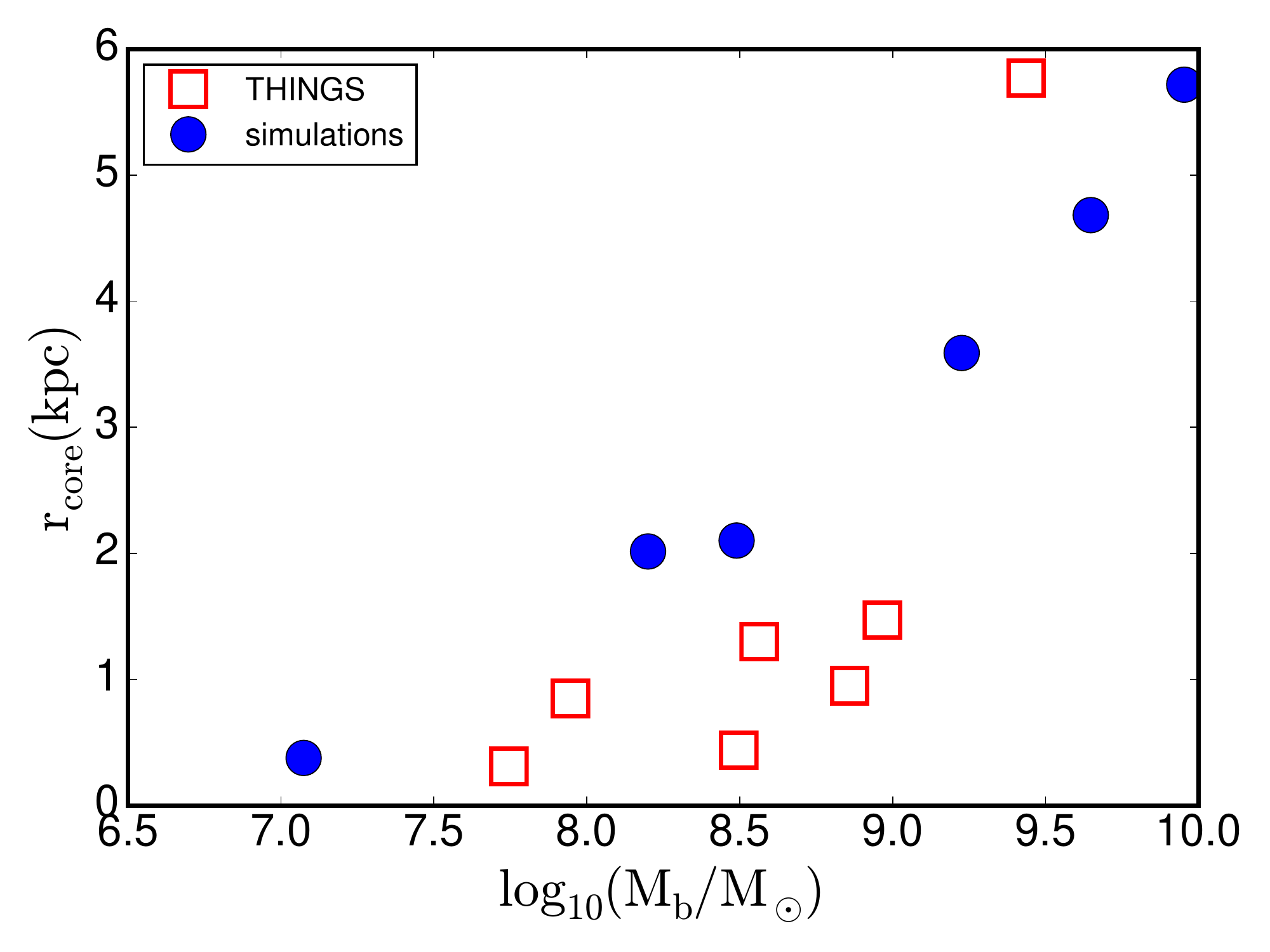}
\caption{Comparison between the core radii of THINGS dwarf galaxies and our simulations as a function of baryonic mass $m_{\rm b}$ (see footnote \ref{mbdef}) at $z=0$. Red hollow squares show the observed core radii \citep{THINGS,Oh11} while blue circles show the core radii from our simulations. Within the plot range, the core size increases with baryonic mass, which is largely consistent with the observed cores. A larger number of observed and simulated dwarfs is necessary to draw stronger conclusions.} 
\label{THINGS_rcore}
\end{figure}

Figure \ref{THINGS_rcore} shows the DM core radii of our simulated galaxies as a function of their baryonic masses and compares them to observations from \cite{THINGS}. In general, for $m_{\rm b} \sim 10^7-10^{10} M_\odot$ \footnote{ \label{mbdef} $m_{\rm b}$ is defined as the total baryonic mass within 20\% $R_{\rm vir}$ of the halo.} , core radius increases with baryonic mass. There is a broad agreement with observed core sizes. A more detailed comparison will require a larger number of halos and more detailed modeling of the methodology used to calculate the core sizes in observed galaxies.

\section{Discussion}
\label{discussion}
We use the FIRE suite of high-resolution cosmological ``zoom-in'' simulations with physical feedback models to study the properties of the central regions of dark matter halos.
Our simulations have higher resolution (in most cases the highest resolution to date at a given halo mass), and more explicit and comprehensive implementation of stellar feedback than simulations used previously to study baryonic effects on DM profiles. Critically our models contain no ``freely adjusted'' parameters tuned to any particular results. We characterize the evolution of halo properties and correlation with galaxy and halo mass and explore the effects of star formation driven feedback and adiabatic contraction on the slopes, cores size and circular velocity profiles of galactic halos.

We find that at $z=0$ the central slopes of DM density profiles measured at $\sim 1 \% \rvir$ are shallow in halos $M_{\rm h} \sim 10^{10}-10^{11}\msun$, but the slopes are cuspier at lower and higher masses. 
We see a sharp transition around $M_{\rm h} \sim 10^{10}\msun$ from cuspier profiles at lower masses to shallow, core-like profiles at higher masses. 
Efficient feedback continues to at least $M_{\rm h} \sim 10^{12}\msun$ where the core-like profiles that form during earlier evolution are contracted by baryons at late times into steeper profiles. Final profiles are similar or flatter than NFW, and therefore significantly flatter than expectations for a contracted NFW halo.

Our results are in broad agreement with others found in the literature. For example we find that feedback is efficient in forming large cores at halo masses $\sim 10^{10}\msun$- few $\times 10^{11}\msun$ which is similar to previous findings \citep{Gove12, DiCi14,Pont14}. It appears that simulations with bursty star formation and outflows, but different small scale feedback implementations, affect dark matter profiles in a qualitatively similar way.  However we also found some interesting differences. This is not surprising, given that our simulations are in most cases higher resolution and have explicit treatment of feedback on small scales. For example we find that in the halos with large cores, cores are established at around $z \sim 1$ and can grow even at low redshift while other authors find that cores formed very early already at $z=2-3$ \citep[e.g.][]{Mada14}. This could be a consequence of different star formation histories that can change core formation \citep{Onor15}. For example, these authors note that earlier simulations or dwarf galaxies with “sub-grid” feedback models produced too many stars at a given halo mass. Alternatively, these differences could indicate that different treatments of material affected by feedback (cooling prevention vs. explicit model) cause differences in DM profiles.
 
We also find cuspier DM profiles at around $6-7 \times 10^{11}\msun$ than the one reported in \cite{Macc12b} and therefore weaker halo expansion and somewhat steeper dependence of slope on $M_*/M_{\rm h}$ than the relation presented in \cite{DiCi14}. Overall, better statistics from a larger number of simulated halos are needed for a more robust analysis of these differences.

\subsection{Energetics}

Given that the efficiency of conversion of halo gas into stars increases from dwarf galaxies to massive galaxies, it is natural to connect feedback effects to the energy available from stellar feedback. We therefore compare the amount of energy available from feedback with the energy needed to overcome the gravitational potential and move dark matter outside of the central region. The simplest estimate is to calculate how much energy gets injected into the ISM from the SNe only, which represents a lower limit to the available energy budget.

\begin{table}
    \centering
    \begin{tabular}{llll}

    \hline\hline
    ~    & $\Delta U\,(1\%\, {\rm core})$& $\Delta U\,(3\%\, {\rm core})$& $E_{\rm sn}$      \\ 
    ~	&(erg)	&(erg)	&(erg) \\ \hline\hline
    {\bf m09} &5.85e52  & 6.45e53 & 2.68e53  \\
    {\bf m10} &1.08e53  & 2.02e54 & 2.13e55 \\
    {\bf m11} &3.43e55  & 6.46e56 & 2.22e58   \\ \hline
    \end{tabular}
    \caption{Comparison of the difference in gravitational potential energy $\Delta U$ of halos from dark matter only simulation and the same halos with constant density core of 1\% of $R_{\rm vir}$ (left) or 3\% of $R_{\rm vir}$ (middle), and their total supernova energy ($E_{\rm sn}$, right), obtained from the corresponding simulations with feedback. Energy input from supernovae alone is sufficient to produce cores in {\bf m10} and {\bf m11}.}
    \label{artcore}
\end{table}

Previously, \cite{Gned02}, \cite{Ogiy11} and \cite{Garr13} claimed that the total feedback energy released in SNe is insufficient to remove enough dark matter to form large cores. In particular, \cite{Garr13} tested if stellar feedback can lower central densities to a degree needed to explain the ``Too Big To Fail'' problem. They simulated supernova feedback with time-varying potential and found the number of SNe needed to match observed profile of a halo hosting $M_*\sim 10^{6}\msun$ galaxy exceeds the number of SNe produced in most of the dwarf galaxies for the typical initial mass function.  However, they did not consider the full growth history of the halo. The frequent mergers and star formation bursts at higher redshift could dynamically  heat up more dark matter in the center, when halo was smaller. Furthermore, they assume mass loading of SNe driven winds smaller than we find for FIRE galaxies of comparable masses \citep{Mura15}. Finally their selected halo has a high concentration compared to a typical sub-halo that is expected to host observed dwarf galaxies. As we show below, there is a sufficient amount of energy available to couple to dark matter in the relevant halo and galaxy mass regime.

Table \ref{artcore} shows the energy needed to create a constant density core with radius $1\%$ or $3\% R_{\rm vir}$ in halos from our dark matter only simulations and the total supernova energy inferred from our hydrodynamic simulations by z=0. We constructed a constant density core in our ``cuspy'' DM only simulations by keeping the density profile outside the core unchanged and moving the excess mass within the core radius to infinity. Then we calculate the total potential energy for the initial and cored profiles with the formula below
\begin{equation}
U=-4\pi G\int_0^{R_{\rm vir}}M(r)\rho(r)r\mathrm{d}r,
\end{equation}   
and report the difference, $\Delta U$, in Table \ref{artcore}.

To estimate the supernova energy, we assume the energy from one supernova $E_{\rm sn}=10^{51}$erg, the fraction of massive stars which can produce supernovae $\xi(m_*>8\msun)=0.0037$ \citep{Krou02}, and the mean stellar mass $M_{\rm mean}=0.4\msun$. The total supernova energy of a halo is given by 
\begin{equation}
E^{\rm tot}_{\rm SN}=M_*/(<m_*>)*\xi(m_*>8\msun)*E_{\rm sn}.
\end{equation}

From Table \ref{artcore} we see that for {\bf m09}, the amount of SNe energy is not sufficient to create a large core at 3\% $R_{\rm vir}$ even if it all couples (via secondary gravitational interactions) to the DM. Even creating 1\% $R_{\rm vir}$ core is difficult as it requires that more than 20\% of the available energy is coupled to dark matter, which is unlikely given the indirect connection via the change of gravitational potential and that a large fraction of this energy is in heavily mass loaded winds that move rapidly \citep{Mura15}.

However, for {\bf m10}, a small core is energetically possible and even a 3\%$R_{\rm vir}$ core requires less than 10\% of the available energy.  We see some signs of profile flattening within the inner 1\% of $R_{\rm vir}$, but not a fully developed core. However, for the same halo mass \cite{Onor15} show that a slightly different star formation history can cause a much larger effect and form a central core with a radius of 1-2 $\% R_{\rm vir}$. 

In {\bf m11}, the supernova energy is three orders of magnitude higher than what is needed to create a small core at 1 \% of $R_{\rm vir}$ and even 3\% of $R_{\rm vir}$ core can be created with few percents of coupling efficiency. It is therefore not surprising that this halo indeed hosts a large core. Even though the depth of the potential well increases in more massive halos, the amount of stellar mass is a steep function of halo mass, which provides sufficient energy to affect central dark matter profiles. This is why we see a relatively sharp transition at $\sim 10^{10}\msun$ from low mass cuspy halos to core-like halos at higher masses.

\subsection{Correlation with star formation}
\label{cwsf}
\label{masslosssfrcor}

Here we show a simple connection between bursts of star formation that cause strong gas ejection episodes, and the change in the amount of dark matter in the central part of halos. In Figure \ref{correlmasslosssfr}, we plot the change in the enclosed dark matter mass as a function of the peak star formation rate in {\bf m11}, the halo in which the effect of feedback on dark matter is one of the strongest in our sample. We focus on bursts with high peaks of star formation and neglect low star formation rate episodes as they typically do not show strong outflows \citep{Mura15}. In Figure \ref{correlmasslosssfr} we plot the rate of change in the central amount of dark matter as a function of the peak star formation rate. The star formation rate and mass change are measured as averages over 0.2 Gyr. The rate of change of the DM mass is 
\begin{equation}
\frac{\Delta M}{\Delta t}(t)=\frac{m_{\rm dm}(t+0.2\rm{Gyr})-m_{\rm dm}(t)}{0.2 \rm{Gyr}},
\end{equation}
where $m_{\rm dm}(t)$ is the enclosed DM mass within $2\% R_{\rm v0}$ (the virial radius at $z=0$)  averaged over 0.2Gyr and t is the beginning time of the interval. We measure star formation between $t-0.2$Gyr and $t$, within a larger radius of 0.1 $R_{\rm v0}$ to make sure to include stars that move out of the very center between the two time intervals. 
\footnote {We have also tried to average SFR over 0.1 Gyr and considered the delay of 0.1 Gyr. The enclosed DM mass drops 0.1 Gyr after the peak SFR in a way similar to Figure 12 (i.e. it is negative), but with the larger scatter and greater changes in DM mass. We note that this simple measurement cannot use much longer intervals for averaging of SF and for time delay because on large time scales the SFR bursts would be “washed out” and we could have multiple burst episodes within a long delay time interval.}

A significant decrease of mass enclosed in the central ${2\%R_{\rm v0}}$ occurs just after a strong starburst when a large amount of dark matter is ``heated'' and effectively pushed out of the central region. At the corresponding times DM only simulation does not show negative mass change, except in the case of the largest burst which is triggered by the close passage during a merger (as indicated in Figure \ref{Rviralphaf}), i.e. the merger dynamically alters the profile. Hence we conclude that there is a correlation between mass removal from the center and star formation rate.

\begin{figure}
\includegraphics[width=0.5\textwidth,natwidth=810,natheight=842]{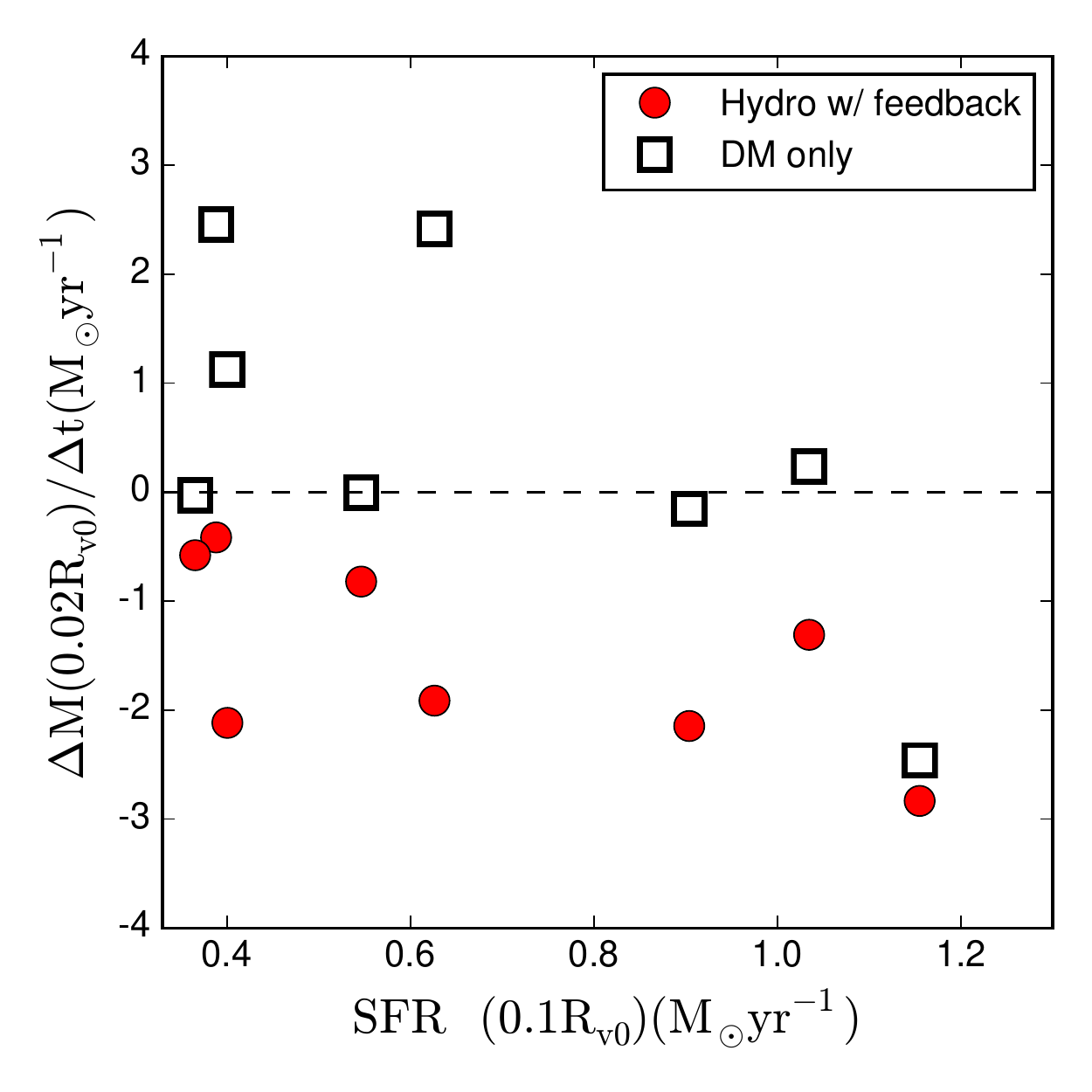}
\caption{The rate of change of enclosed dark matter mass within $2\%$ of $R_{\rm v0}$ as a function of peak star formation rate within $10\% R_{\rm v0}$ in {\bf m11} (red circles) and {\bf dm11} (black hollow squares) simulations, both averaged over 0.2 Gyr. In both cases we plot the SFR from the corresponding feedback simulations at the same cosmic time. The star formation rate is measured 0.2 Gyr ahead of the rate of change of enclosed dark matter mass. Strong bursts of star formation are followed by a reduction in the enclosed DM mass in the simulation with feedback.}
\label{correlmasslosssfr}
\end{figure}

This correlation suggests that a strong burst of star formation can provide sufficient feedback to remove a significant amount of baryons, which then cause a decrease in the central potential and lower the concentration of DM. This scenario is consistent with the mechanism suggested in \cite{Pont12} in which repeated changes of central gravitational potential transfer energy to the orbits of DM particles causing a central density decrease. 

Some fraction of the removed DM does return, i.e. cores get partially rebuilt, however when large cores are present this is a relatively small effect.  When cores are established after the period of rapid early halo growth, they survive largely intact for at least several Gyrs, before a new major bursts of star formation and gas expulsion sets in (see \S ~\ref{timevo}). In more massive halos ({\bf m12}), shallow profiles that form at higher redshift steepen by redshift zero significantly. In these halos we see continuous late time star formation without significant galactic wind episodes \citep{Mura15} which continuously increases central density of baryons that dominate the central potential. As a consequence dark matter halos are pulled inward by adiabatic contraction changing shallow profiles into cusps.

While a burst-driven core formation mechanism is consistent with our findings, we cannot exclude a contribution of other dynamical mechanisms, such as the motion of dense baryonic clumps within galaxies with respect to the halo centers. On the other hand, core formation via enhanced dynamical friction from the dense infalling sub-structure \citep[e.g.][]{ElZa01} is unlikely to play a significant role for the halo mass range probed here, because feedback lowers the density of infalling sub-halos relative to their DM-only counterparts (see Figure \ref{densityf}; all of the infalling sub-halos in our sample have $M_{\rm h}\lesssim 10^{11}\msun$).

\subsection{Significance for the dark matter detection}

The dark matter profile in the Milk Way is important in studies of indirect detection of dark matter particles from annihilation or decay signals. Recently, extended emission in gamma rays from the galactic center has been reported based on the data from the Large Area Telescope aboard the Fermi Gamma Ray Space Telescope \citep[e.g.][]{Hoop11a}. To interpret this signal as a consequence of annihilation of dark matter particles or to constrain its contribution it would be extremely useful to know the central dark matter profile \citep{Abaz14}.
The signal is consistent with the annihilation rate of thermally produced weakly interacting massive particle dark matter \citep{Good09,Hoop11a,Hoop11b,Boya11,Abaz12,Gord13,Maci14,Abaz14} although other possibilities remain valid alternatives \citep{Abaz11}.

While we have a small number of objects at the relevant mass, one robust finding from our simulations is that central density of cold dark matter will not correspond to contracted NFW profiles in Milky Way-mass halos. This helps constrain the range of values used in the modeling of the observed signatures. We find values of  $\alpha \sim -1$ to $-1.4$ at $1-2\% \rvir$  which is consistent with the best fit in \cite{Abaz14}. However our results from {\bf m12} halos suggests that deeper in the halo, the DM profile is likely shallower $\alpha \sim -0.7 \; \mathrm{to} -1.1$ at $< 1\% \rvir$. Given the slight differences of the central slope definition and fitting procedures it would be interesting to test if the profiles shown here provide a good match for the observed signal. We defer such more detailed comparison for future work.

\subsection{Limitations}
\label{limitations}

There are some clear limitations in our study of dark matter halo properties. The number of simulated halos in our sample is limited because such high-resolution simulations are very time-consuming. In the future we plan to simulate a much larger number of halos to be able to extract direct statistical predictions and to compare to observed values, including the scatter in observations and theoretical predictions. 

Further limitation comes from the finite resolution. Our simulations are amongst the highest resolution cosmological simulations at z=0 to date and our {\bf m09 and m10} simulations can robustly resolve dark matter profiles on $< 200$ pc scales. However detailed comparison to observations requires  converged results and the ability to exactly integrate circular orbits over a Hubble time without N-body effects within $300$ pc in $M_{\rm h} \sim 10^{10-11}\msun$($M_* \sim 10^{7-9}\msun$) halos \citep{THINGS}. In our {\bf m10h1297}, {\bf m10h1146} and {\bf m11} simulations, this is larger than many gravitational softening lengths, however the more stringent Power convergence criterion  \S \ref{resolimit} \citep{Powe03} requires a further factor of several improvement in mass resolution to reach this limit.

Resolution requirements can be even more dramatic if one wants to directly study the inner density profiles of small dwarfs that form within sub-halos of large spiral galaxies (e.g. to directly address TBTF problem). Such study would require simulating $M_{\rm h}=10^{12}\msun$ halos with $> 10^9$ particles within the virial radius. We have shown, however, that many of the CDM ``problems'' can be explained by baryonic feedback in isolated, well resolved, dwarf galaxies suggesting a similar solution for satellite galaxies.

We caution that the Power convergence criterion was derived for different time-stepping algorithm, different force accuracy and different softening values. Based on the DM profiles in Figure \ref{densitynobc} we see that steep central slope in DM only simulations continue to at least factor of $\sim 2$ smaller radii than what is estimated by Power convergence radius. More importantly, direct resolution tests confirm the convergence in DM cusp profiles down to a factor $\sim 2-3$ smaller radii for our fiducial resolution. This suggests that this convergence criterion might be too conservative for our simulation setup.

We did not study {\bf m13} halo ($M_{\rm h}\sim 10^{13}\msun$) from the FIRE project sample because the star formation rate of this halo, at low redshift, is higher than the observed rates in observed galaxies of the same mass (H14). Some additional physics, e.g. active galactic nucleus (AGN) feedback, is needed to explain the observations. Yet, the investigation of the profiles of these high mass halos is highly intriguing both observationally and theoretically. Observationally, gravitational lensing provides an accurate measurement of the enclosed mass of those halos \citep{Bolt08}, though dark matter only constitutes a minor fraction of the mass.

Nevertheless, stellar kinematics and strong lensing do suggest cores in galaxy clusters with $\alpha \sim -0.5$ \citep{Newm13a,Newm13b}. It is not clear if a single mechanism can explain such shallow slopes over a large range. Mechanisms ranging from more frequent major mergers as well as processes such as AGN feedback, magnetic fields, anisotropic conduction, and cosmic rays, that are not yet incorporated in our simulations may be important in regulating late time star formation and affecting the core formation in massive halos  \citep[e.g.][]{Peir08,Mart13}.  

AGN feedback in particular might even affect halos with masses similar to our {\bf m12} simulations, e.g. \cite{Vell14} showed that AGN feedback can have a non-negligible impact on the halo properties (i.e. mass and profile) down to $M_{\rm h} \sim 5\times 10^{11}\msun$. These results suggest that the effect of AGN feedback, in addition to stellar feedback, could further lower the central density of the most massive halos in our sample. We note that regardless of the dominant feedback mechanism, the overall efficiency of feedback must be similar to what is seen in our simulations, as this efficiency is constrained by the observed $M_*-M_{\rm h}$ relation. Our simulations provide a clean test for the effects of stellar feedback alone on the DM distribution.

\subsection{Dependence on star formation history}

\cite{Onor15} compared our {\bf m10} simulation from H14 with the one with a slightly different supernova feedback coupling at smaller scales. In our default case, energy deposition is volume weighted while in the other version it was mass weighted. This creates slight differences in the feedback and changes late time star formation. A 1-kpc core was formed in a halo with more prominent late time star formation, while our default {\bf m10} simulation shows a core at $< 400$pc and much higher central density of dark matter. It is likely that this strong sensitivity is caused by this halo mass being at the transition region from smaller to larger cores.

This then suggests that when a large sample of simulated halos is available, comparison to the observations around this mass scale will potentially help distinguish between feedback models by analyzing their star formation histories and properties of their dark matter halos (Fitts et al., in preparation). For slightly more massive halos we see core formations in all cases we explored, regardless of their detailed SF history and the details of their feedback implementations.

\section{Conclusions}
We have explored cold dark matter profiles in simulations with stellar feedback. We used the FIRE suite of hydrodynamic simulations initially discussed in H14 and supplement these with 4 new dwarf galaxy simulations. We also run collisionless counterparts for all of these simulations. We show that baryonic simulations can successfully produce results consistent with observations and alleviate or solve several so called ``problems'' of the CDM: ``cusp-core'', the lack of adiabatic contraction or ``halo expansion'' and the ``Too Big To Fail'', without any fine turning or introduction of adjustable parameters. Our main results are:

1. The baryons have little influence on halos with  $M_{\rm h}\ll 10^{10}\msun$ because only a small fraction of available baryons are converted to stars, owing to feedback and the UV background that suppress their star formation after the reionization \citep{Onor15, Whee15}. The smallest halos are therefore perfect places for testing various theories of dark matter.

2. The central slopes of dark matter density profiles are governed by halo mass and stellar mass. Profiles are shallow with relatively large cores for $M_{\rm h} \sim 10^{10} -$ few $ \times 10^{11}\msun$ and  $M_* \sim 10^7- $ few $ \times 10^9 \msun$, where $\alpha \sim -0.5-0$ and cores are $r_{\rm core} > 1$ kpc. Small central cores can also form at slightly lower masses $M_* \sim 10^6\msun$. This result is consistent with the observations of dwarf galaxies and can explain the ``cusp-core" problem.

3. Bursts of star formation and feedback start forming cores at early times but the cores are established typically at later times, e.g. in our {\bf m11} simulation core is still growing at $z<1$. Stable cores are established once central regions of halos stop their rapid growth. After this time ($z\lesssim 2$) removal of mass from the central region leaves a long term effect on the halo profile. We show that strong bursts of star formation are correlated with dark matter expansion.

4. The total supernova energy in halos with $M_{\rm h}>10^{10}\msun$ is sufficient to produce a core with radius $3\% R_{\rm vir}$, but not sufficient to make large cores in lower mass halos. In practice only a few percent of the available energy is transferred into evacuation of dark matter from the central region.

5. Baryonic contraction of dark matter halos becomes significant when central regions of halos are clearly dominated by baryons, which in our simulations occurs for $ M_{\rm h} > 5\times 10^{11} \msun$. However feedback in the progenitors of these massive galaxies significantly lowered central DM density at $z\sim 1-1.5$. The cumulative effect of feedback and contraction is then a profile that in our {\bf m12} runs is slightly shallower than NFW. This explains why the normalization of the Tully-Fisher relation requires no contraction or halo expansion with respect to the collisionless NFW profile.

6. Stellar feedback in galaxies with $M_* \sim$ few $\times 10^6-10^8\msun$ lowers the central density of DM when compared to dark matter only simulations and significantly reduces the rotational velocity near the center. This means that relatively low circular velocities of observed galaxies should correspond to much higher maximum circular velocities or virial velocities of halos or sub-halos in collisionless CDM simulation. This can solve or at least substantially alleviate the ``Too Big To Fail" problem.

\section{ACKNOWLEDGEMETS}

We would like to thank N. Murray, J. Bullock, M. Boylan-Kolchin and P. Salucci for useful discussion.
DK and TKC were supported in part by NSF grant AST-1412153, \rm{Hellman Fellowship} and funds from the University of California San Diego. Support for PFH was provided by an Alfred P. Sloan Research Fellowship, NASA ATP Grant NNX14AH35G, and NSF Collaborative Research Grant \#1411920. EQ was supported in part by NASA ATP grant 12-APT12-0183, a Simons Investigator award from the Simons Foundation, and the David and Lucile Packard Foundation. CAFG was supported by NSF through grant AST-1412836, by NASA through grant NNX15AB22G, and by Northwestern University funds. The simulation presented here used computational resources granted by the Extreme Science and Engineering Discovery Environment (XSEDE), which is supported by National Science Foundation grant number OCI-1053575, specifically allocations TG-AST120025 (PI Kere\v{s}), TG-AST130039 (PI Hopkins), TG-AST1140023 (PI Faucher-Gigu\`ere).

\bibliographystyle{mn2e}
\bibliography{mn-jour,mybib}
\appendix
\section{Convergence limits}
\label{relimit}
We use the Power convergence criterion \citep{Powe03} to derive empirical formulae for the minimum particle mass needed to quantify cusps down to $0.2-3\% R_{\rm vir}$. We consider DM only simulations and assume their profiles can be fitted with an NFW profile. The ratio between the scale radius $r_{\rm s}$ and the virial radius $R_{\rm vir}$ is determined through a concentration-mass relation from \cite{Dutt14}, and we use the virial overdensity from \cite{Brya98}. Then we calculate the enclosed number of particles and density within $0.3-3\% R_{\rm vir}$ as well as 300 and 700 pc. From Eq. \ref{relimeq}, we estimate the minimum radius such that $t_{\rm relax}/t_0<0.6$ and plot the relation between the required particle mass to meet this criteria at the desired radius, and the total halo mass in Figure \ref{relimfig}.
\begin{figure}
\includegraphics[scale=0.40]{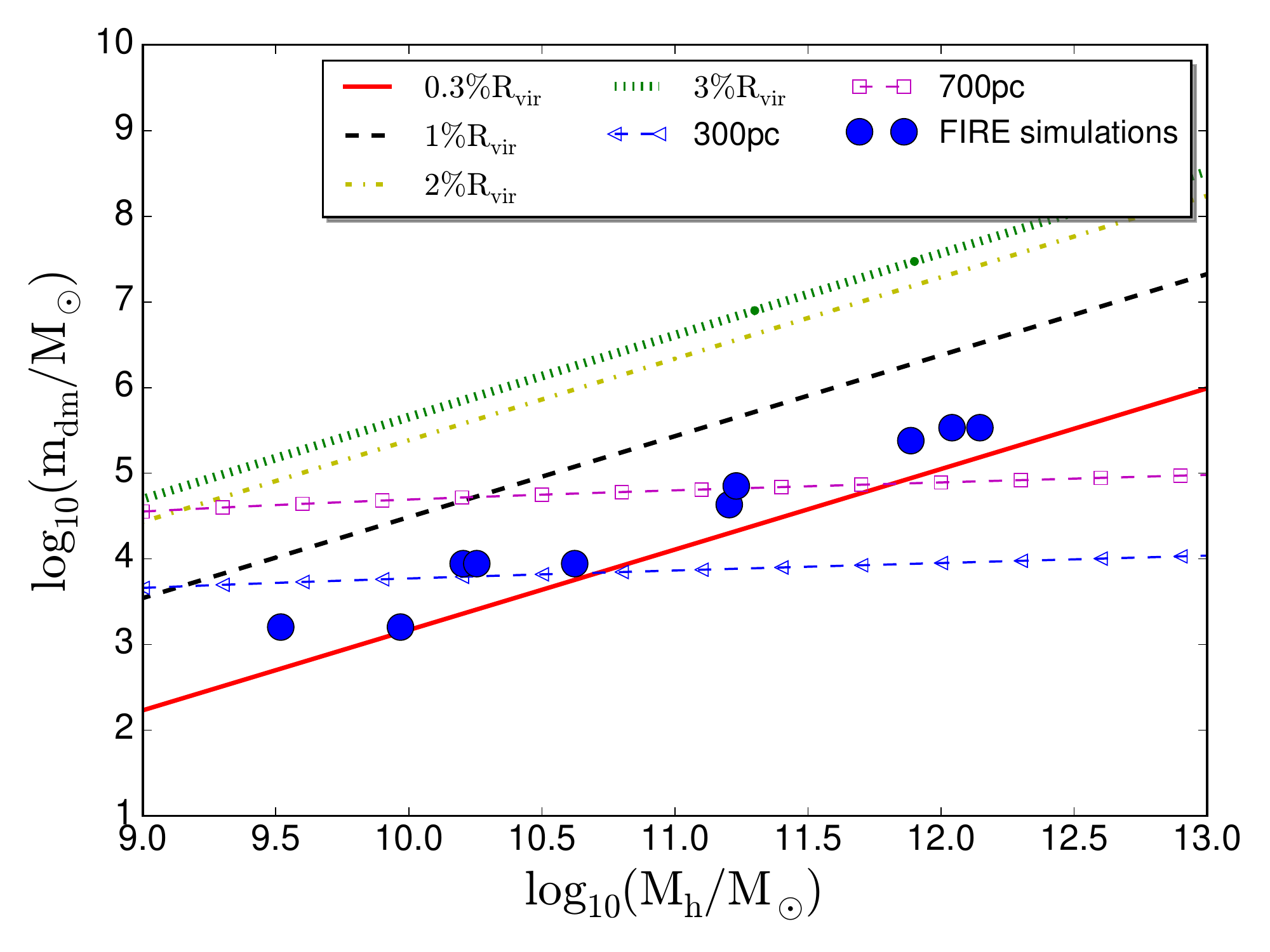}
\caption{The virial mass of a halo as a function of the maximum particle mass allowed in order to reach convergent profiles at $0.3,1,2, 3\% R_{\rm vir}$, 300 pc and 700 pc. Small circles represent the corresponding values from dark matter only simulation listed in Table \ref{SIC}. (Our full physics simulations have the same DM particle numbers.)}
\label{relimfig}
\end{figure}

The data can be fitted with a linear relation,
\begin{equation}
\log_{10}(m_{\rm dm}/\msun)=a\log_{10}(M_{\rm h}/\msun)+b
\label{resoeq}
\end{equation}
where $a$ and $b$ are list in Table \ref{relimeqtab}
\begin{table}
	\centering
    \begin{tabular}{lll}
    \hline\hline
    $r_{\rm res}/R_{\rm vir}$           & $a$           & $b$               \\ \hline\hline
    $0.3\%$                     & 0.94          & -6.2               \\
    $1\%$                       & 0.95          & -5.0               \\
    $2\%$                       & 0.95          & -4.3               \\ 
    $3\%$                       & 0.96          & -3.9               \\
    \hline
    \end{tabular}
    \caption{The coefficients in Eq. \eqref{resoeq} for different resolutions.}
    \label{relimeqtab}
\end{table}

\cite{Oh11} measured $\alpha$ between 300-700 pc in dwarf galaxies with $M_*$ between $10^6-10^9M_\odot$ and found cored profiles in those galaxies. In order to match this observational result, the minimum resolved radius in the simulations should be around 300 pc. For a fixed physical radius this turns into a requirement in particle mass that is almost independent of the halo mass, owing to higher concentration in lower mass halos.The maximum mass of dark matter particles needs to be slightly smaller than $10^{4}M_\odot$ to converge at 300 pc and smaller than $10^5 M_\odot$ to converge within 700 pc. Our {\bf m09} and {\bf m10} are clearly converged at both of these radii.

Our slightly more massive halos {\bf m10h1297}, {\bf m10h1146} and {\bf m10h573} are marginally converged at 300 pc, but fully converged at 700 pc. Figure \ref {relimfig} assumes an NFW profile and implicitly assumes that the central region has close to a Hubble time to undergo relaxation processes. As discussed in the main text, it is not clear what the appropriate convergence criterion should be once large cores are formed and the central density is reduced. This likely depends on the core formation time as well as details of the gravitational softening of multiple particle species and their time-stepping algorithms.

It is important to note that our DM force softening is typically a factor of five smaller than the corresponding Power convergence criterion. Furthermore, we have also tested if the force softening of the baryonic component influences dark matter profiles: e.g. we increased the baryonic softening from 2.0 to 25 pc in the slightly modified version of {\bf m10} run in \cite{Onor15} and found that the dark matter profile was only mildly changed (the core size was actually larger in run with smaller softening). While two-body relaxation effects are important in estimating central DM profiles, in H14 (Appendix C) we have used idealized runs to show that our standard resolution in {\bf m12} runs is also sufficient to reliably determine other relevant quantities, e.g. SFR, wind mass-loading and gas phase distribution. All are consistent to within a factor of $\sim 2$ even with $\sim 50$ times better mass resolution. This indicates that the general properties of dark matter profiles on resolved scales in our simulations are numerically robust.

\section{Choice of $\alpha$}
\label{choiceofalpha}
We investigate the effect of different fitting ranges on $\alpha$ in this appendix. 
We consider three different fitting ranges, 1-2 kpc, 0.3-0.7 kpc, and $1-2\% R_{\rm v0}$. Figure \ref{Mvir_alpha} shows $\alpha$ that corresponds to those ranges. 
\begin{figure}
\includegraphics[scale=0.4]{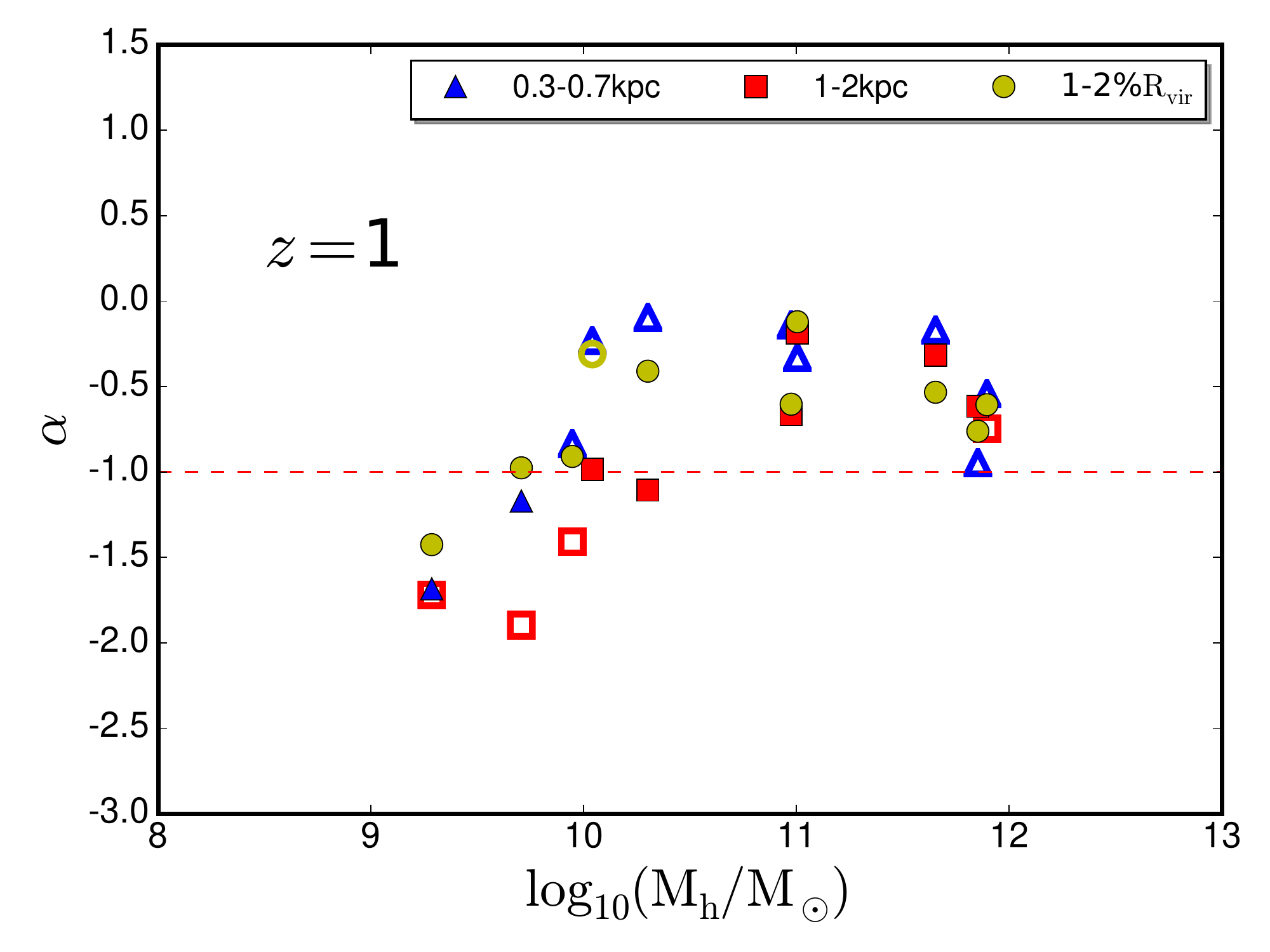}
\includegraphics[scale=0.4]{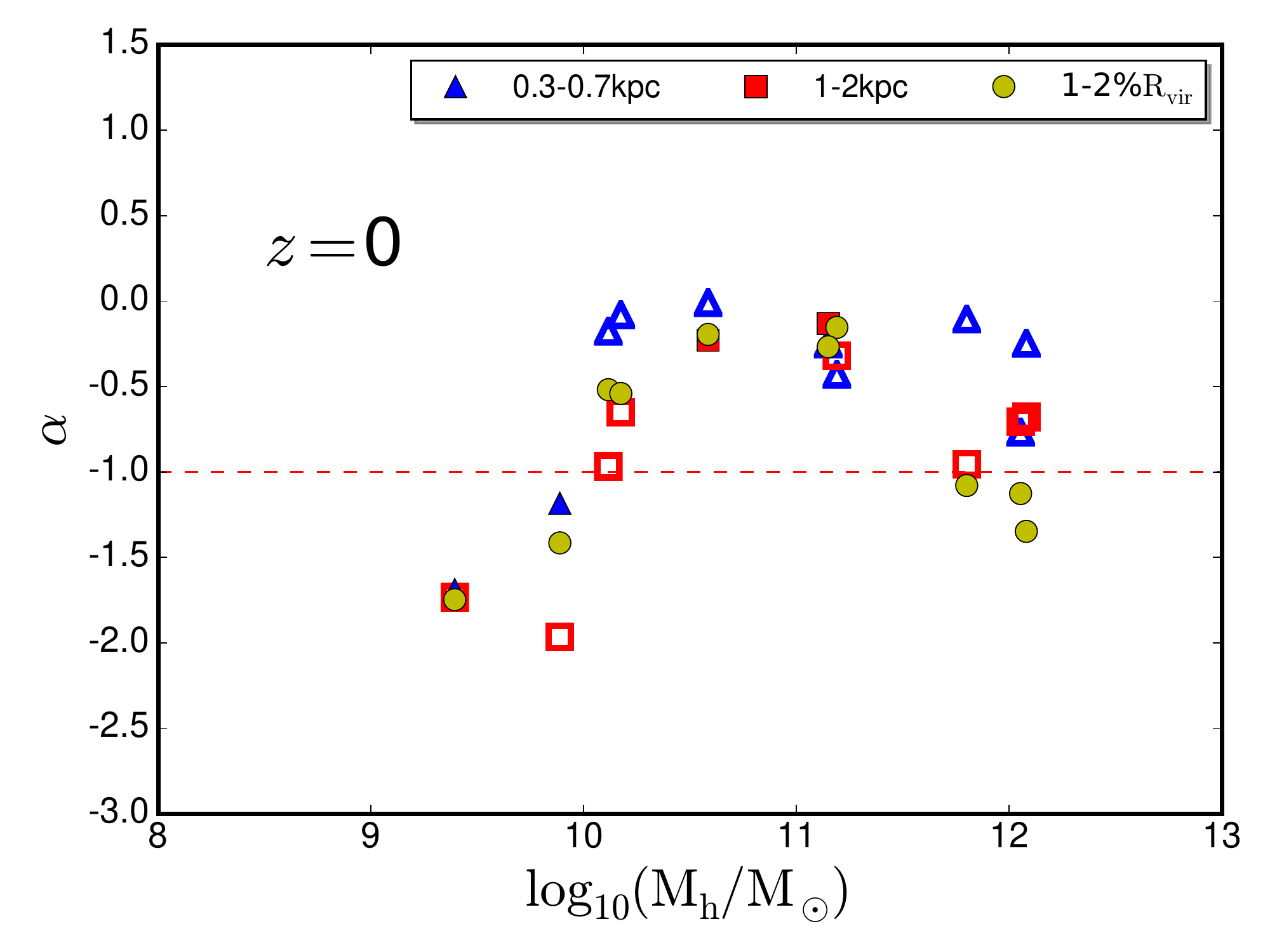}
\caption{DM profile slope $\alpha$ inferred from different criteria plotted as a function of the halo mass. ``0.3-0.7kpc'' is the dark matter density slope interpolated within 0.3-0.7 kpc from the center. ``$1-2\% R_{\rm vir}$'' is $\alpha$ interpolated within $1-2\% R_{\rm vir}$.  ``1-2kpc'' is $\alpha$ interpolated within 1-2 kpc. Filled circles show that the profile measurement range is larger than 0.5$r_{\rm Pow}$ and smaller than a third of $r_{\rm s}$. Open circles indicate that one of these criteria is not satisfied.}
\label{Mvir_alpha}
\end{figure}
In general, 0.3-0.7 kpc includes some overlap below the Power radius for halos with mass larger than $10^{11}\msun$ but $\alpha$ in this range can be directly compared with observations of dwarf galaxies. 1-2 kpc lies outside the central region ($>1.3 r_{\rm s}$) in small dwarfs. $1-2\% R_{\rm vir}$ is well-resolved, lies inside the central region and physically meaningful, so we use this fitting range as our ``default'' choice in the main text. Overall all of the methods show very similar trends of the DM density profile slope with mass.

\bsp

\label{lastpage}

\end{document}